\documentclass[fleqn,usenatbib]{mnras}

\usepackage{newtxtext,newtxmath}

\usepackage[T1]{fontenc}
\usepackage{subfig}
\DeclareRobustCommand{\VAN}[3]{#2}
\let\VANthebibliography\thebibliography
\def\thebibliography{\DeclareRobustCommand{\VAN}[3]{##3}\VANthebibliography}

\usepackage{graphicx}	
\usepackage{amsmath}	
\usepackage{pdflscape}

\title[Multiple-arc model of the Plutinos]{A new multiple-arc model of the resonant Kuiper belt objects -- Plutinos}

\author[Chen \& Li]{
Yue Chen$^{1,2}$
and Jian Li$^{1,2}$\thanks{E-mail: ljian@nju.edu.cn}
\\
$^{1}$School of Astronomy and Space Science, Nanjing University, 163 Xianlin Avenue, Nanjing 210023, PR China\\
$^{2}$Key Laboratory of Modern Astronomy and Astrophysics in Ministry of Education, Nanjing University, Nanjing 210023, PR China
}
\date{Accepted XXX. Received YYY; in original form ZZZ}

\pubyear{2015}

\begin{document}
\label{firstpage}
\pagerange{\pageref{firstpage}--\pageref{lastpage}}
\maketitle

\begin{abstract}
To incorporate the gravitational influence of Kuiper belt objects (KBOs) in planetary ephemerides, uniform-ring models are commonly employed. In this paper, for representing the KBO population residing in Neptune's 2:3 mean motion resonance (MMR), known as the Plutinos, we introduce a three-arc model by considering their resonant characteristics. Each `arc' refers to a segment of the uniform ring and comprises an appropriate number of point masses. Then the total perturbation of Plutinos is numerically measured by the change in the Sun-Neptune distance ($\Delta d_{SN}$). We conduct a comprehensive investigation to take into account various azimuthal and radial distributions associated with the resonant amplitudes ($A$) and eccentricities ($e$) of Plutinos, respectively. The results show that over a 100-year period: (1) at the smallest $e=0.05$, the Sun-Neptune distance change $\Delta d_{SN}$ caused by Plutinos decreases significantly as $A$ reduces. It can deviate from the value of $\Delta d_{SN}$ obtained in the ring model by approximately 100 km; (2) as $e$ increases in the medium range of 0.1-0.2, the difference in $\Delta d_{SN}$ between the arc and ring models becomes increasingly significant; (3) at the largest $e\gtrsim0.25$, $\Delta d_{SN}$ can approach zero regardless of $A$, and the arc and ring models exhibit a substantial difference in $\Delta d_{SN}$, reaching up to 170 km. Then the applicability of our three-arc model is further verified by comparing it to the perturbations induced by observed Plutinos on the positions of both Neptune and Saturn. Moreover, the concept of the multiple-arc model, designed for Plutinos, can be easily extended to other MMRs densely populated by small bodies. 

\end{abstract}




\section{Introduction}

\label{sec:intro}

The accuracy of planetary ephemerides is crucial for space exploration, and therefore, continuous development efforts are underway \citep[e.g.][]{newhall1983102, pitjeva2001modern, fienga2008inpop06, pitjeva2014development, folkner2014planetary, fienga2020inpop}. Traditionally, the ephemeris computation includes mutual perturbations of the Sun, the eight planets, the Pluto system, and the satellites, along with considerations for the solar oblateness J2, the solar pressure, figure and tide effects, relativistic corrections, lunar librations, and so on. \citep{pitjeva2014development}. In addition to the influencing factors mentioned above, modern planetary ephemerides have been updated to incorporate the gravitational perturbations of a vast number of asteroids observed in the solar system \citep{folkner2014planetary, fienga2008inpop06, pitjeva2014development, Tian2023}. The contributions mainly come from three asteroid populations: main belt asteroids, Jupiter Trojans, and Kuiper belt objects.

The main belt asteroids (MBAs) are distributed between the orbits of Mars and Jupiter, with a current estimated population of around 700,000 \citep{li2019calibration}. To simulate the total perturbation caused by the MBAs, the `Bigs + ring' model is commonly adopted. The `Bigs' refer to the MBAs that have the most significant influence on the Earth-Mars distance (e.g. 1 Ceres, 2 Pallas), and they are individually included in the numerical ephemerides calculations \citep{Kuchynka2010}. Generally, ephemerides contain about 300 `Bigs', for which only the self-gravity of the three largest asteroids (i.e. 1 Ceres, 2 Pallas, and 4 Vesta) is considered \citep{pitjeva2018masses, fienga2020inpop} to speed up the calculation. As for the `ring', which is homogeneous, it represents the remaining numerous MBAs and is simply implemented as a gravitational potential \citep{KRASINSKY200298, fienga2008inpop06, pitjeva2014development}. In the ephemeris INPOP06, \citet{fienga2008inpop06} estimated that the perturbation induced by this asteroid ring on the Earth-Mars distance can reach up to 150 m during the time interval of 1969-2010. To reduce the time cost in ephemeris calculations, \citet{liu2022ring} proposed an alternative approach known as the multiple-ring model. They took into account all MBAs, regardless of size, and classified them into 122 families based on their proper elements \citep{Nesv2015}. Eventually, they assigned the MBAs to six rings with varying heliocentric radii and inclinations. With regard to the total perturbation of the MBAs, the six-ring model yields a Mars-Earth distance that deviates from the value obtained in the `Bigs + ring' model by an error below 0.5 m over a 10-yr period. This error is acceptable, given that tracking data from Mars spacecraft are available with metric precision \citep{Kuchynka2010}. The main advantage of this new approach is that it improves computational efficiency, as there is no need to incorporate a large quantity of individual MBAs (i.e. the `Bigs') in the numerical ephemerides.

Jupiter Trojans (JTs) are asteroids that share the orbit of Jupiter, but leading and trailing Jupiter by about $60^{\circ}$ in longitude, i.e. around the L4 and L5 triangular Lagrangian points. These asteroids are said to be settled in the 1:1 mean motion resonance with Jupiter. As of February 2023, more than 12,000 JTs are registered in the Minor Planet Center (MPC) \footnote{https://minorplanetcenter.net/iau/lists/MPLists.html}. Since the L4 and L5 swarms are distributed in two separate regions, respectively, this unique configuration of JTs can not be simulated by a simple ring model like those used for the MBAs. Furthermore, there are many more JTs in the L4 swarm than the L5 swarm, with a number ratio expected to be between 1.3 and 2 \citep{jewi04, szabo2007properties, grav2011wise, grav2012wise}. According to \citet{li2023JT}, this number asymmetry holds for JTs with absolute magnitudes $H<15$ (equivalent to diameters $D>7$ km, assuming a typical albedo of 0.04), while the fainter objects are too few to be statistically significant. The number asymmetry of JTs also contradicts the assumption of a homogeneous-ring model. In \citet{li2018constructing}, besides the 226 largest JTs with $H<11$, the authors modeled the remaining objects with $H>11$ using two arcs located around Jupiter's L4 and L5 points, respectively. Their findings demonstrated that the total effect of JTs can lead to a change of $\sim70$ m in the Earth-Mars distance during the 2014-2114 time interval.

The Kuiper belt objects (KBOs) are icy celestial bodies located beyond the orbit of Neptune. The orbits of KBOs can be grouped into three classes \citep{gladman2008}: (1) Resonant KBOs: These objects occupy the mean motion resonances (MMRs) with Neptune, and they have small to large eccentricities ranging from $e=0.05$ to 0.35. A significant portion of resonant KBOs is found in Neptune's 2:3 MMR at $\sim39.4$ AU, sharing this resonance with Pluto. This particular population is called `Plutinos', and \citet{alexandersen2016carefully} estimated that there are about 9000 Plutinos with diameters $D> 100$ km. Numerous resonant KBOs are also located in Neptune's 1:2 MMR at $\sim47.8$ AU, comprising an estimated population of about 4400 objects with $D> 100$ km \citep{chenYT2019}. It is worth noting that Neptune's 2:3 MMR and 1:2 MMR positions are usually considered to be the inner and outer boundaries for the main Kuiper belt \citep{petit2023hot}. In addition, the Outer Solar System Origins Survey (OSSOS) has enriched other MMRs. For example, the 2:5 resonant KBOs are more numerous than initially anticipated, and their total population may be as large as that of the Plutinos \citep{volk2016ossos}. (2) Classical KBOs: These objects are not in Neptune's MMRs and typically exhibit small to moderate eccentricities and inclinations. Approximately 30000 non-resonant KBOs with $D> 100$ km are estimated to exist in the main Kuiper belt \citep{petit2023hot}. (3) Scattered KBOs: These objects have highly eccentric orbits, with perihelion distances $q>30$ AU and  semi-major axes $a>50$ AU \citep{FERNANDEZ2004372,jewitt1998large}. \citet{lawler2018ossos} proposed that there are about 90000 scattered KBOs with $D> 100$ km.

Since the late 2000s, the dynamical model of KBOs in planetary ephemerides has undergone gradual refinement. When modeling the Kuiper belt, most of the previous ephemerides mainly focused on the main Kuiper belt \citep{fienga2022evolution,pitjeva2018mass,pitjeva2018masses}. In the ephemeris EPM2008, besides the 21 biggest KBOs, the perturbations caused by the remaining smaller objects were modeled by a one-dimensional ring with a heliocentric radius of 43 AU \citep{pitjeva_2010}. By adjusting the mass of the ring to fit the observation data obtained from spacecraft, Pitjeva estimated that the total mass of KBOs should not exceed $0.027M_{\oplus}$, where $M_{\oplus}$ denotes the Earth mass. In the updated ephemeris EPM2013, \citet{pitjeva2014development} improved the KBO model by individually considering the 31 most massive objects, while the smaller objects were still collectively represented by a single ring. Considering that the inner and outer edges of the main Kuiper belt correspond to Neptune's 2:3 and 1:2 MMRs, respectively, \citet{pitjeva2018mass} introduced an 8 AU-wide annulus spanning from 39.4 AU to 47.8 AU (i.e. a two-dimensional ring) to represent the total perturbation of numerous small KBOs. However, such a two-dimensional ring model poses a severe drawback as objects with different heliocentric distance at the annulus rotate with the same angular velocity. This is obviously different from the reality. To overcome this drawback, these authors developed a new model comprising three separate rings: two rings positioned at 39.4 AU and 47.8AU represent the 2:3 and 1:2 resonant KBOs, respectively, and the third ring is placed at 44 AU, symbolizing the `core' of the Kuiper belt predominantly inhabited by the classical KBOs. In a recent study, \citet{di2020analysis} also employed the three-ring model to estimate the mass of the Kuiper belt. By fitting the high-precision measurements of Saturn obtained from the \textit{Cassini} mission, they provided a total mass for the Kuiper belt of $0.061 \pm 0.001M_{\oplus}$.

When dealing with a large number of uniformly distributed asteroids, a single- or multiple-ring model would be appropriate for representing their total perturbation. However, for the resonant KBOs, their motions are restricted to specific regions within the resonance's phase space, rather than encompassing it entirely \citep{li2022machine}, i.e. they may not distributed uniformly. It is evident that Neptune's MMRs lead to special distributions of these KBOs' phases with respect to Neptune, and this factor can affect the perturbation induced by the resonant KBOs. Accordingly, we aim to construct a new model to represent the total perturbation of these resonators, and estimate the level of the inaccuracy introduced when using a ring model. 

 \begin{table*}
  \caption{Eleven of the most massive KBOs currently known.}
  \label{table:1} 
  \begin{tabular}{cccc}
    \hline
Number & Name & Mass $(10^{-4}M_{\oplus}$) & Reference\\
\hline 
136199 & Eris & $27.96\pm0.33$ & \citet{brown2007mass}\\
134340 & Pluto + Charon & $24.47\pm0.11$ & \citet{brozovic2015orbits}\\
136108 & Haumea & $6.708\pm0.067$ & \citet{ragozzine2009orbits}\\
136472 & Makemake & $4.35\pm0.84$ & \citet{pitjeva2018mass}\\
225088 & 2007 OR10 & $2.93\pm0.117$ & \citet{kiss2019mass}\\
50000 & Quaoar & $1.67\pm0.17$ & \citet{fraser2013mass}\\
90482 & Orcus & $1.0589\pm0.0084$ & \citet{brown2010size}\\
208996 & 2003 AZ84 & $0.69\pm0.33$ & \citet{pitjeva2018mass}\\
120347 & Salacia & $0.733\pm0.027$ & \citet{stansberry2012physical}\\
174567 & Varda & $0.446\pm0.011$ & \citet{grundy2015mutual}\\
55637 & 2022 UX25 & $0.2093\pm0.0050$ & \citet{brown2013density}\\
    \hline
  \end{tabular}
 \end{table*}

In this paper, we focus on the Plutinos, which constitute the largest resonant population observed in the Kuiper belt. The resonant angle $\sigma$ associated with Neptune's 2:3 MMR is defined by
\begin {equation} 
\sigma=3\lambda-2\lambda_N-\varpi,
\label{eq:resonantargument}
\end {equation}
where $\lambda$ and $\varpi$ are the mean longitude and the longitude of perihelion of the Plutino, respectively, and $\lambda_N$ is the mean longitude of Neptune. For a stable Plutino, its resonant angle $\sigma$ librates around $180^{\circ}$ with a resonant amplitude of $A<180^{\circ}$. Here we designate the resonant amplitude as a half-width of the variation of $\sigma$. Consequently, $\sigma$ cannot traverse from 0 to $360^{\circ}$, indicating that the mean longitudes of Plutinos may not evenly spread across the range of 0-$360^{\circ}$. The primary objective of this study is to develop a dynamical model that account for this inhomogeneity in the azimuthal distribution of Plutinos.

In our previous study on the perturbation of JTs \citep{li2018constructing}, a similar issue regarding the azimuthal distribution was discussed. In that study, JTs are allowed to be on circular orbits with eccentricities $e=0$ because the resonant term of the 1:1 Jovian MMR in the expansion of the disturbing function does not contain $e$ (see chap. 6 of \citet{murray_dermott_2000}). However, Neptune's 2:3 resonance is the eccentricity-type, and its strength is proportional to $e$. Consequently, Plutinos must possess $e$ values greater than 0. When involving the contribution of $e$, their orbital configuration becomes more complex. Thus it is necessary to additionally consider the resulting asymmetry in the radial distribution of Plutinos.

The rest of this paper is organised as follows. In Sect. \ref{sec:dym}, we design the dynamical model of the solar system and choose the change in the Sun-Neptune distance as a measurement for the perturbation of the KBOs, including the Plutinos. In Sect. \ref{sec:perturbation}, we construct both the ring model and arc model to simulate the total perturbation of the Plutinos, followed by a comparison of the results obtained from these two models. We conduct a detailed investigation of the plausible azimuthal and radial distributions of the arcs, which are determined by Plutinos' resonant amplitudes $A$ and eccentricities $e$, respectively. We then provide a concise analytic expression to describe the contributions of $A$, $e$, and the total mass of Plutinos to the change in the Sun-Neptune distance. In Sect. \ref{sec:real}, we verify the applicability of our three-arc model by comparing it to the perturbations induced by observed Plutinos with debiased orbits on the positions of both Neptune and Saturn. The conclusions and discussion are given in Sect. \ref{sec:conclu}. 

\section{Dynamical model of the solar system}
\label{sec:dym}

Before studying the Plutinos, this section initiates the construction of our dynamical model, focusing on the entire KBO population. The unperturbed model of the solar system comprises the Sun and eight planets from Mercury to Neptune. The planets’ masses, initial heliocentric positions, and velocities are taken from DE405 \citep{standish1998jpl}. First, their orbits are adjusted from the mean equatorial system to the J2000.0 heliocentric ecliptic system at epoch 2021 July 5. From this epoch forward, we construct the perturbed model of the solar system by incorporating the gravitational perturbations of massive KBOs. In the subsequent analysis, our goal is to assess the influence of KBOs on the modern planetary ephemerides. To achieve this, we compare the motion of Neptune, the closest planet to the KBOs, in both the unperturbed and perturbed models.

To quantify the gravitational effect of KBOs on the orbit of Neptune, we examine the change in the distance between the Sun and Neptune, defined as
\begin{equation}
 \Delta d_{SN}=d_{SN1}-d_{SN0},
 \label{dsn}
\end{equation}
where $d_{SN1}$ and $d_{SN0}$ indicate the Sun-Neptune distances calculated with and without perturbations from the KBOs, respectively. We note that although the planetary ephemerides should be parameterized in the barycentric coordinate system, it is common to assess the validity of a perturbation model using the relative distance between two celestial bodies (e.g. Sun-planet, planet-planet), rather than the barycentric distance \citep{Kuchynka2010, Pitjeva2013, li2018constructing, pitjeva2018mass, liu2022ring}. This approach is preferred due to the variation of the solar system barycentre (SSB) in different perturbation models. For instance, the inclusion of all the asteroids in the main belt and Kuiper belt can cause a displacement of the SSB on the order of 100 km \citep{li2019calibration}. Consequently, the barycentric positions of the Sun and all other celestial bodies would change significantly, but the relative coordinates remain the same.

In our numerical simulations of the solar system's evolution, we employ the 19th-order Cowell prediction-correction algorithm (PECE) with a time-step of 0.5 days, which is chosen based on the orbital period of the innermost celestial body (i.e. Mercury) in our models \citep{tian1993adams, li2018constructing}; and the data output step is set to 5 days. In this N-body code we take into account gravitational interactions among the Sun, planets, and KBOs but ignore the effects of KBO self-gravity.

\begin{figure}
    \includegraphics[width=\columnwidth]{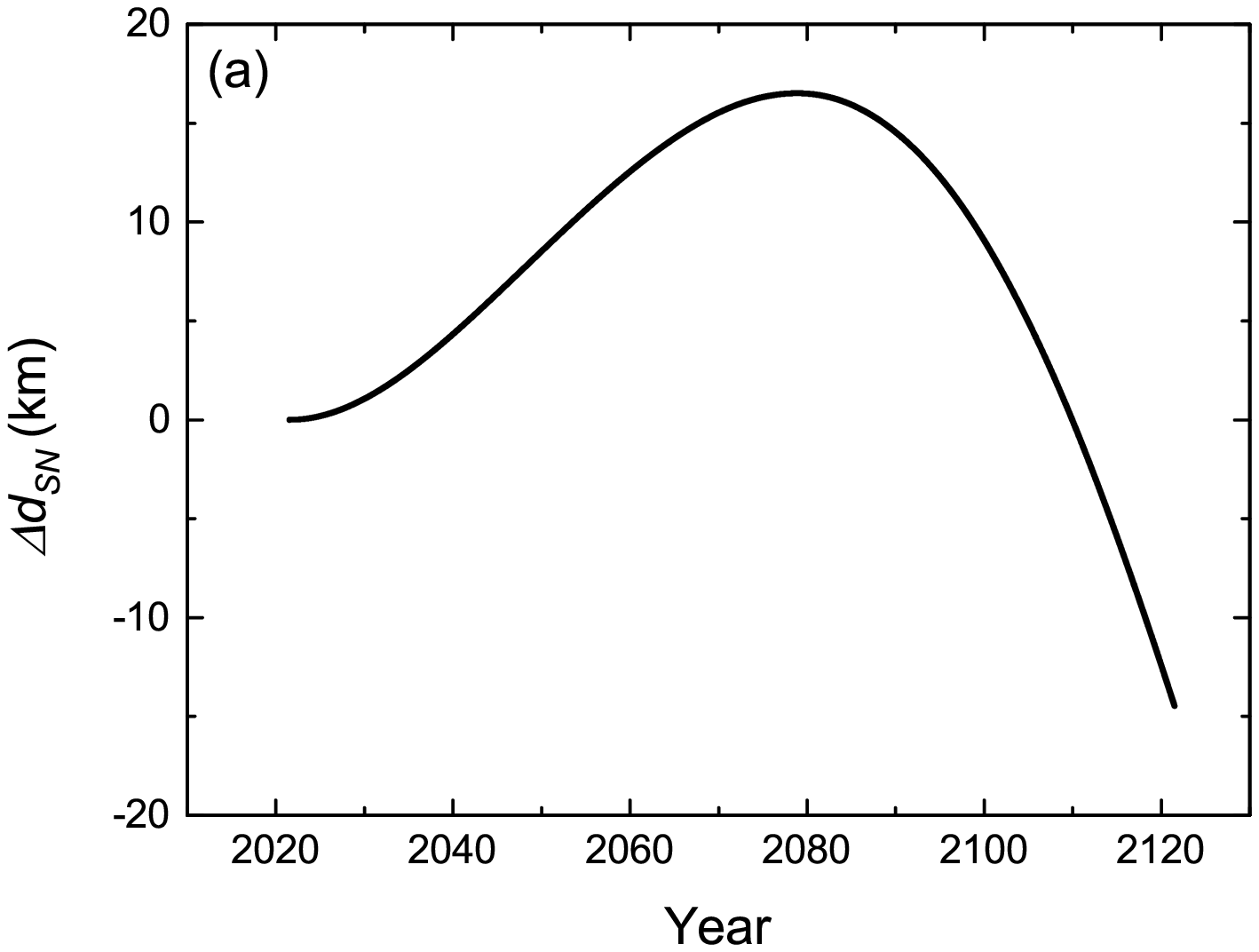}
    \includegraphics[width=\columnwidth]{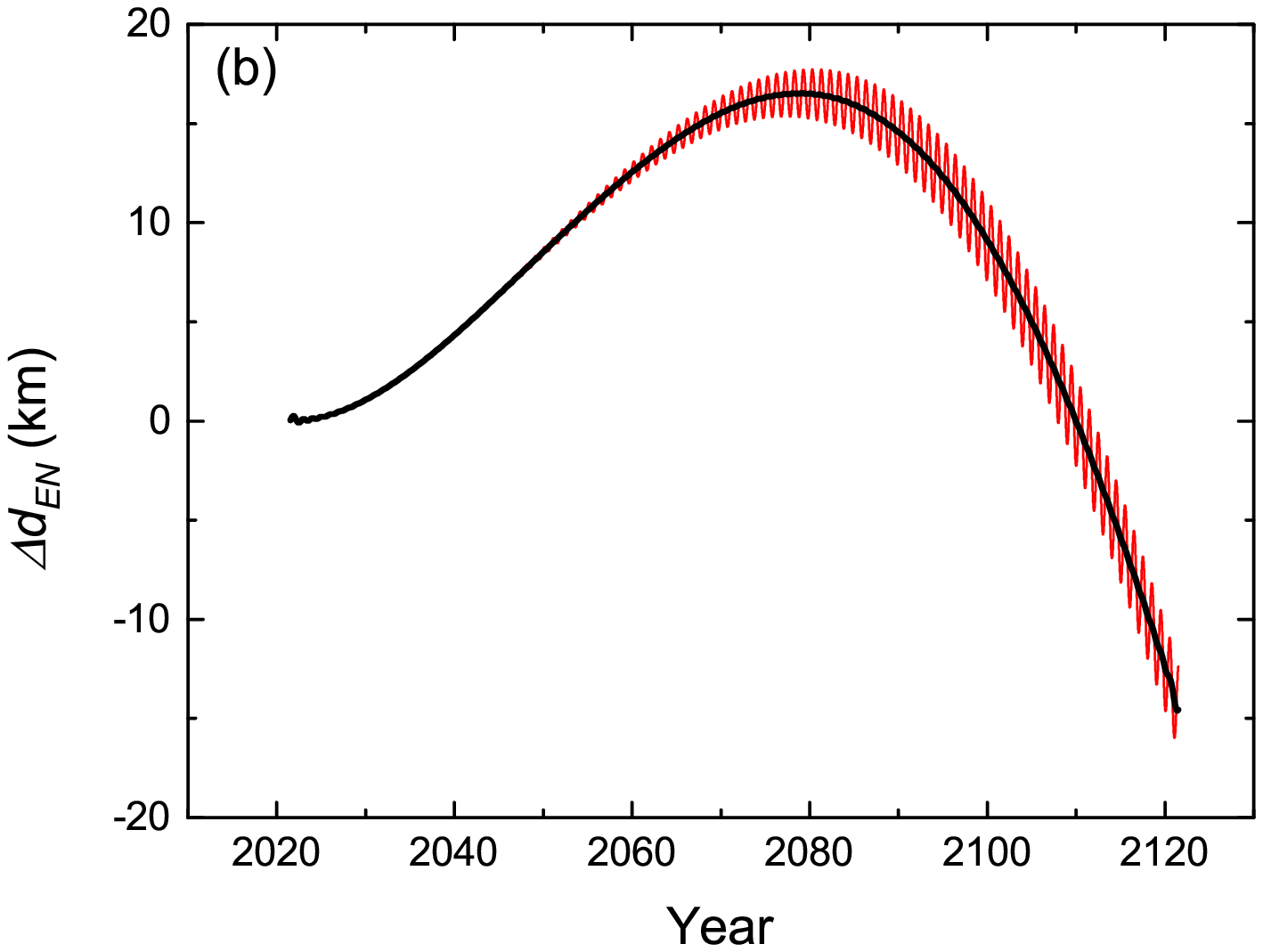}
    \caption{Perturbation on the position of Neptune induced by the 11 most massive KBOs in the time interval between the years 2021 and 2121: (a) the change in the Sun-Neptune distance, denoted as $\Delta d_{SN}$; (b) the change in the Earth-Neptune distance, denoted as $\Delta d_{EN}$. In this panel, the red curve represents the intrinsic value of $\Delta d_{EN}$, while the black curve depicts the secular behaviour of $\Delta d_{EN}$.}
    \label{fig:1}
\end{figure}

First of all, we begin by estimating the magnitude of the perturbation caused by the KBO population on the Sun-Neptune distance. For simplicity, we here consider the influence of the 11 most massive KBOs that have their individual satellites, as listed in Table \ref{table:1}. From the motion of their satellites, their masses are well determined by applying Kepler's third law. The orbital elements of these objects are gathered from the MPC at the specific epoch of 2021 July 5, as mentioned earlier. This way, we can measure the perturbations from these 11 prominent KBOs through the change in the Sun-Neptune distance, denoted as $\Delta d_{SN}$. Figure \ref{fig:1}(a) shows that, between the years 2021 and 2121, the resultant $\Delta d_{SN}$ could attain an approximate value of 16.5 km. This substantial value indicates a significant perturbation caused by the KBOs on the Sun-Neptune distance. As a result, modern planetary ephemerides may have to incorporate the gravitational effects of the KBOs.

It is important to note that the 11 selected objects represent only a fraction of the entire KBO population. These objects collectively possess a total mass of about $0.007 M_{\oplus}$. While the total mass of KBOs could be an order of magnitude larger, reaching up to $0.2M_{\oplus}$ as estimated theoretically by \citet{pitjeva2018mass}. Therefore, it is reasonable to suppose that the combined influence of all KBOs on Neptune's position could be even stronger. Moreover, the spatial distribution of KBOs will also play a significant role, as we will discuss later when examining the case of Plutinos.

In this paper, we will use the Sun-Neptune distance to measure the perturbations of KBOs. However, given that the current observations are typically conducted from Earth or its vicinity, it is more customary to determine a planet's position relative to Earth \citep{standish1998jpl,folkner2014planetary,fienga2020inpop}. For example, the Earth-Mars distance has been commonly used to measure perturbations induced by the MBAs \citep{fienga2008inpop06} or the JTs \citep{li2018constructing}. To validate our choice of using the change in the Sun-Neptune distance, with or without considering the 11 massive KBOs listed in Table \ref{table:1}, we also calculate the change in the Earth-Neptune distance:
\begin{equation}
 \Delta d_{EN}=d_{EN1}-d_{EN0},
\end{equation}
where the variables have the similar meanings to Eq. (\ref{dsn}), but the subscript `E' indicates Earth. In Fig. \ref{fig:1}(b), the red curve illustrates the temporal variation of $\Delta d_{EN}$ resulting from the perturbation of these 11 KBOs. Over a period of 100 yr, the change in the Earth-Neptune distance can reach a value as large as $\Delta d_{EN}\sim17.7$ km, comparable to the largest $\Delta d_{SN}$ displayed in Fig. \ref{fig:1}(a).

Upon examining the red $\Delta d_{EN}$-curve in Fig. \ref{fig:1}(b), one can observe short-period oscillations on a timescale of 1 yr, corresponding to Earth's revolution around the Sun. To extract the low-frequency change in the Earth-Neptune distance, we applied a fast Fourier transform (FFT) low-pass filter to $\Delta d_{EN}$. This process yields the smooth black curve, as plotted in Fig. \ref{fig:1}(b). Comparing the black curves in Figs. \ref{fig:1}(a) and (b), we find a remarkable agreement between $\Delta d_{SN}$ and the smoothed $\Delta d_{EN}$, with a difference between these two quantities being less than 0.4 km. Based on the results obtained here and above, we propose that the Sun-Neptune distance can serve as a representative measurement for evaluating the gravitational perturbations caused by the KBOs.

Adopting the Sun-Neptune distance as a metric offers several advantages. First, it simplifies the description of the secular variation of Neptune's position over time by effectively avoiding the short-period changes observed in the Earth-Neptune distance due to the Earth's annual movement. More importantly, another notable advantage is the substantial reduction in computational cost. By using the Sun-Neptune distance, there is no need to provide Earth's position in locating Neptune. This allows us to remove the four terrestrial planets from the dynamical model of the solar system, with Jupiter becoming the innermost planet instead of Mercury. Given the Jupiter-to-Mercury period ratio of $\sim20$, we can increase the integration time-step from 0.5 days to 10 days \citep{li2007origin}. As a result, the numerical integration process would be substantially faster.

In the simplified model of the solar system, the gravitational effects of the terrestrial planets are implemented by adding their combined masses to the Sun. However, before proceeding with further investigations, we need to validate that this approximation does not substantially influence the change in the Sun-Neptune distance caused by the perturbations of the KBOs. To assess this, we once again calculate the value of $\Delta d_{SN}$ resulting from the perturbations of the 11 most massive KBOs within the simplified model. The results reveal that the profile of the time evolution of $\Delta d_{SN}$ remains nearly identical to that obtained in the complete model, which includes all eight planets. Indeed, the relative error in the largest value of $\Delta d_{SN}$ is as small as 0.019\%. Therefore, without compromising our main results, the simplified solar system model can reduce the calculation expense to merely 1/20 of the original amount. 

In a brief summary, our dynamical model for subsequent calculations includes the Sun and the four giant planets, optionally incorporating the KBOs. Furthermore, our attention will be directed toward a specific subset of KBOs, namely the Plutinos in the 2:3 MMR with Neptune. To evaluate the gravitational perturbation induced by the Plutinos, we will utilize the change in the Sun-Neptune distance, i.e. $\Delta d_{SN}$. 

\section{Perturbation of the Plutinos}
\label{sec:perturbation}

When considering a large number of KBOs uniformly distributed in azimuth, their total perturbation should manifest as a continuous ring. By the term `continuous', we mean that the acceleration of a body is computed using an asteroid ring potential \citep[see Eq. (8) in][]{fienga2008inpop06}. However, for the sake of computational simplicity, it is common to approximate the perturbation of KBOs with a discrete ring. In contrast to the continuous ring, the discrete one consists of a number of moving point masses. In \citet{pitjeva2018mass} and \citet{di2020analysis}, 40 point masses were adopted to represent the discrete ring. Although this number may seem very small, it remains reasonable since the masses and positions of these point masses can be adjusted to fit observational data. But our current study aims to theoretically investigate the difference between the ring and arc models, so we have to minimize the impact of such discretization. 

From a geometrical perspective, as the number $n$ of point masses increases, the discrete ring approximation becomes increasingly accurate in representing the continuous ring. Similarly, as the arc is a portion of the ring, the discrete arc will also approach the continuous one as the value of $n$ becomes large enough. Determining an appropriate number of point masses within a ring or arc allows us to achieve a balance between computational efficiency and the robustness of our simulations. It's important to note that, unlike a homogeneous ring that remains independent of time, the arcs will rotate with Neptune, meaning their azimuthal positions constantly change with time. Therefore, the use of a discrete arc model offers an advantage: the point masses can naturally revolve alongside Neptune. Another advantage is that, by assigning eccentricities to the orbits of the point masses in the discrete model, we can easily account for the influence of eccentric Plutinos. 

In the following investigation of both the ring and arc models, we exclusively concentrate on the zero-inclination case. Adopting the planar model aligns with previous works that modeled KBOs using the one-dimensional ring \citep{pitjeva_2010} or the two-dimensional annulus \citep{pitjeva2018mass}. Additionally, it is worth noting that Neptune's 2:3 resonance is, in fact, an eccentric-type resonance \citep{li2020study}. The dynamical features of this resonance are primarily determined by the Plutino's eccentricity, while its inclination has limited significance.

\subsection{The ring model}
\label{subsec:ring}

Considering the ring model, we assume that the mass is distributed evenly across its azimuth. For a continuous ring, its gravitational potential exerting at the location $(x',y', z')$ can be described as
\begin{equation}
  \begin{aligned}
	&U(x',y',z',t)=-\int_r\int_0^{2\pi}\frac{G\rho_ar}{dist}drd\phi \\
	&dist=\sqrt{(rcos\phi-x')^2+(rsin\phi-y')^2+z'^2}\label{eq:ring}
  \end{aligned}
\end{equation}
where $G$ is the gravitational constant, $\rho_a$ is the linear density of the ring, and $(r,\phi)$ are the polar coordinates of the ring relative to the Sun. Notably, the two parameters $\rho_a$ and $r$ are assumed to be fixed, resulting in Eq. (\ref{eq:ring}) with only one variable, i.e. $\phi$. Consequently, the integration can be performed using the Romberg algorithm.

To determine the linear density $\rho_a$ of the ring, it is necessary to obtain the total mass of the Plutinos. The total mass of the entire population of KBOs was estimated to be within a wide range of $M_{kb}=0.01-0.2M_\oplus$ \citep{pitjeva2018mass}. Later, through an analysis of high-precision measurements of Saturn from the \textit{Cassini} mission, \citet{di2020analysis} proposed a more specific mass of $M_{kb}=0.061M_\oplus$, with Plutinos accounting for approximately $1/6$ of this mass. Based on these findings, we will assume a total mass of $M_{plu}=0.01M_\oplus$ for the Plutinos. It should be noted that, to a first-order accuracy, the perturbation caused by the Plutinos is linearly proportional to $M_{plu}$ \citep{li2018constructing}. This implies that, after selecting a specific value for $M_{plu}$ in our calculations, we can easily generalize the results for the perturbation of Plutinos when their total mass is refined in the future observations. Further details on this will be discussed in Sect. \ref{result}.

Regarding the radial distance $r$ of the ring, we should set it to be the nominal 2:3 resonance location of 39.4 AU. However, in order to determine the minimum number density required for our discrete-arc model, we need to first compare the difference between a uniform continuous ring and a uniform discrete ring. In this case, choosing 39.4 AU may pose a potential issue. If the point masses are distributed around 39.4 AU, the uniformity of the discrete ring may be disrupted by Neptune's 2:3 resonance. To avoid this issue when comparing the continuous and discrete ring models, we opt for a slightly larger radial distance of $r=43.4$ AU, which is 1.44 times the Neptune's semi-major axis of $a_N=30.1$ AU. At the heliocentric distance of 43.4 AU, objects with eccentricities smaller than 0.1 are not in the 8th or lower order resonances with Neptune \citep{li2023study}.

\begin{figure}
  \includegraphics[width=\columnwidth]{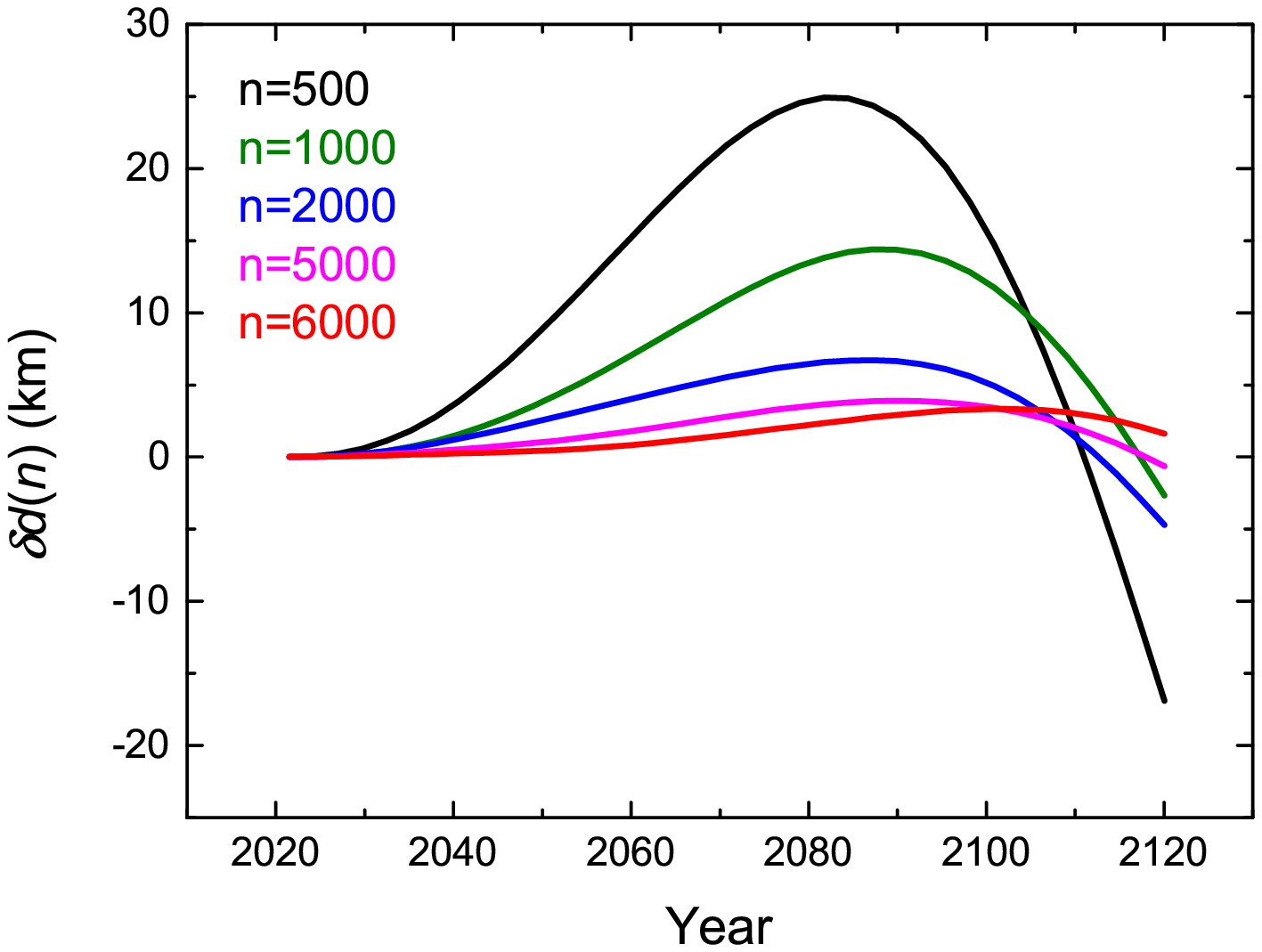}
  \caption{Difference in the change of Sun–Neptune distance, denoted by $\delta d(n)$ (see Eq. (\ref{Dring})), between the continuous and discrete ring models. Each curve indicates the case of a discrete ring consisting of $n=500$ (black), 1000 (green), 2000 (blue), 5000 (magenta), and 6000 (red) point masses.}
  \label{fig:2a}
\end{figure}

\begin{figure}
    \includegraphics[width=\columnwidth]{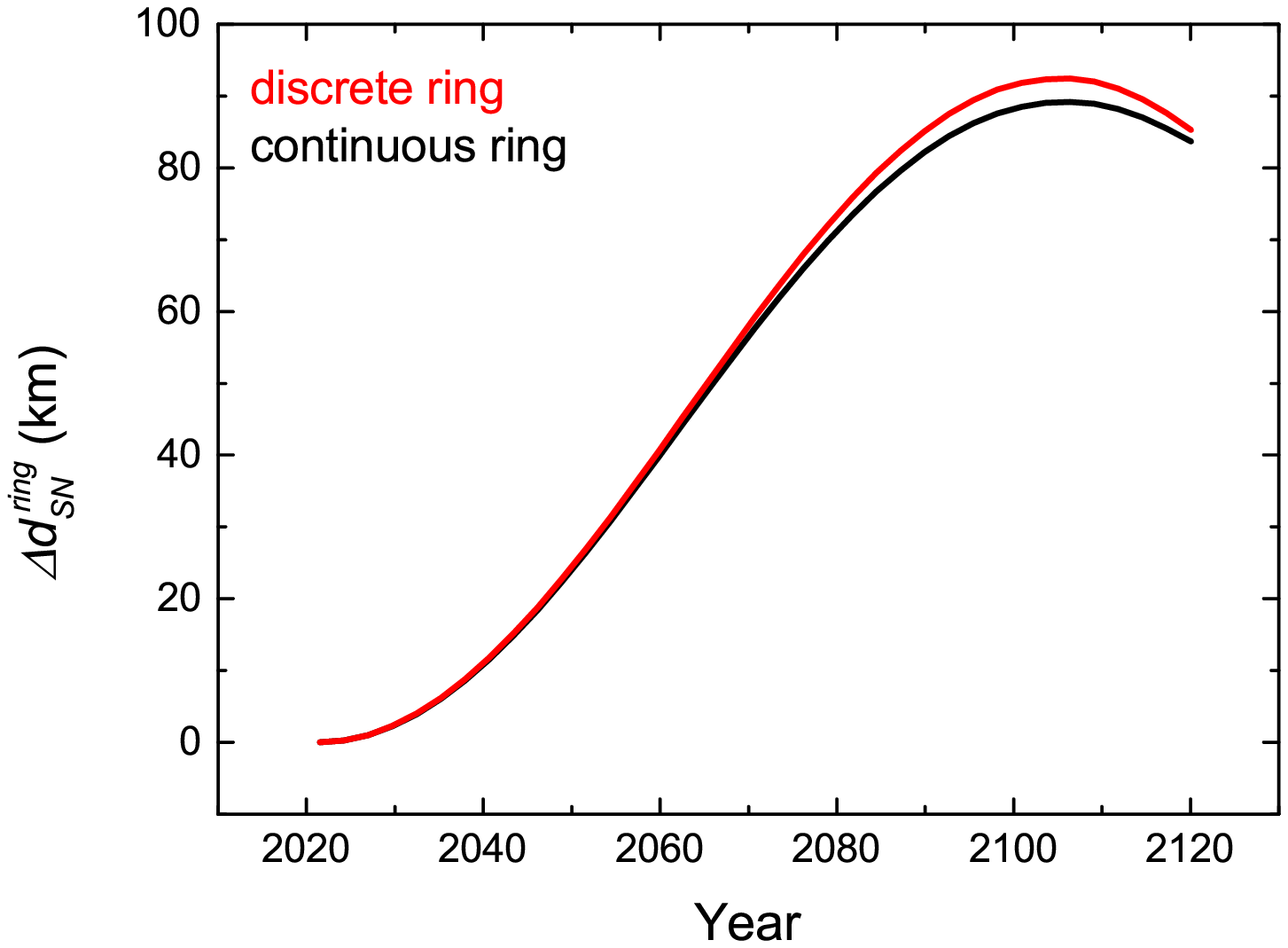}
    \caption{Perturbations on the Sun-Neptune distance induced by the continuous ring (black) and the discrete ring comprised of 6000 point masses (red). This figure corresponds to the specific case of $n=6000$ shown in Fig. \ref{fig:2a}.}
    \label{fig:2b}
\end{figure}

We then calculate the changes in the Sun-Neptune distance induced by both the continuous and discrete rings, denoted as $\Delta d_{SN}^{ring}$ and $\Delta d_{SN}^{ring}(n)$, respectively. The difference between the perturbations of these two rings can be expressed as:
\begin{equation}
  \delta d(n)=\Delta d_{SN}^{ring}(n)-\Delta d_{SN}^{ring},
\label{Dring}
\end{equation}
where $n$ refers to the number of point masses used to construct the discrete ring. As discussed earlier, increasing the value of $n$ helps reduce the effects of discretization, quantified by $\delta d(n)$ in Eq. (\ref{Dring}). We compute the values of $\delta d$ by varying $n$ from 500 to 6000, and the results are shown in Fig. \ref{fig:2a}. It is evident that the difference $\delta d(n)$ exhibits temporal variation due to the evolution of the system. Therefore, we parameterize this difference by its maximum value, denoted as $\max \delta d(n)$, within the considered epoch from 2021 to 2121. For $n=500$, the value of $\max \delta d(n)$ can be as large as about 30 km. However, as $n$ increases to 5000, $\max \delta d(n)$ decreases by an order of magnitude, reaching only 3.8 km. In addition, with a further increase in $n$ to 6000, we observe that $\max \delta d(n)$ decreases by as little as $\sim0.1$ km. Even if we were to continue increasing $n$, it would have a limited impact on decreasing the value of $\max \delta d(n)$. As a matter of fact, $\max \delta d(n)<4$ km measures the absolute error between the continuous and discrete ring models. Considering a value as large as 100 km for $\Delta d_{SN}^{ring}$, the relative error is just about 4\%. 

To better visualize the small discrepancy between the perturbations induced by the continuous and discrete rings, we focus on the case of $n=6000$. Figure \ref{fig:2b} depicts the temporal variations of $\Delta d_{SN}^{ring}$ and $\Delta d_{SN}^{ring}(n)$, represented by the black and red curves, respectively. We observe that the relative error between these two curves remains below 4\% throughout the 100-yr period. Therefore, we conclude that the discrete ring consisting of 6000 point masses is adequate to achieve a reliable approximation of the continuous ring.

By setting $n=6000$ for the discrete ring, its number density $\Sigma_0$ in azimuth can be calculated via 
\begin{equation}
    \Sigma_0=\frac{6000}{2\pi}\frac{r}{r_0},
\label{number_ring}
\end{equation}
where $2\pi$ is the radian measure of a complete ring, $r$ is the heliocentric radius of the ring, and $r_0=43.4$ AU is the reference radius. In the subsequent investigation of the arc model, it is essential to ensure that the number density of the discrete arc satisfies the condition given in Eq. (\ref{number_ring}). Let $l$ be the length of a discrete arc in azimuth, the minimum number of the point masses required to construct this arc can be calculated as:
\begin{equation}
    n_{arc}=\Sigma_0 \times l.
\label{number_arc}
\end{equation}
For the arc-like longitude distribution of Plutinos driven by Neptune's 2:3 resonance, we will first determine the associated length $l$ in the next section.

\subsection{The three-arc model}
\label{subsec:arc}

As the primary focus of this paper, we again note that Plutinos, a distinct group of KBOs trapped in Neptune's 2:3 resonance, are not uniformly distributed. Therefore, modeling them as a homogeneous ring may introduce a certain level of inaccuracy due to their unique spatial distribution. This distribution will be represented by the arc model, in which the Plutinos are arranged into some arcs. By `arc', we are referring to a segment of the ring. We are now about to demonstrate the azimuthal distribution of Plutinos in physical space. For the 2:3 resonance, its critical resonant angle $\sigma$ is defined by Eq. (\ref{eq:resonantargument}). According to the expression for $\sigma$, the azimuthal distribution of Plutinos can be represented by the differences in the mean longitude between the Plutinos and Neptune, expressed as: 
\begin{equation}
\Delta\lambda=\lambda-\lambda_N=\frac{1}{3}(\sigma+\varpi-\lambda_N).
\label{longitude}
\end{equation}

\begin{figure}
  \includegraphics[width=\columnwidth]{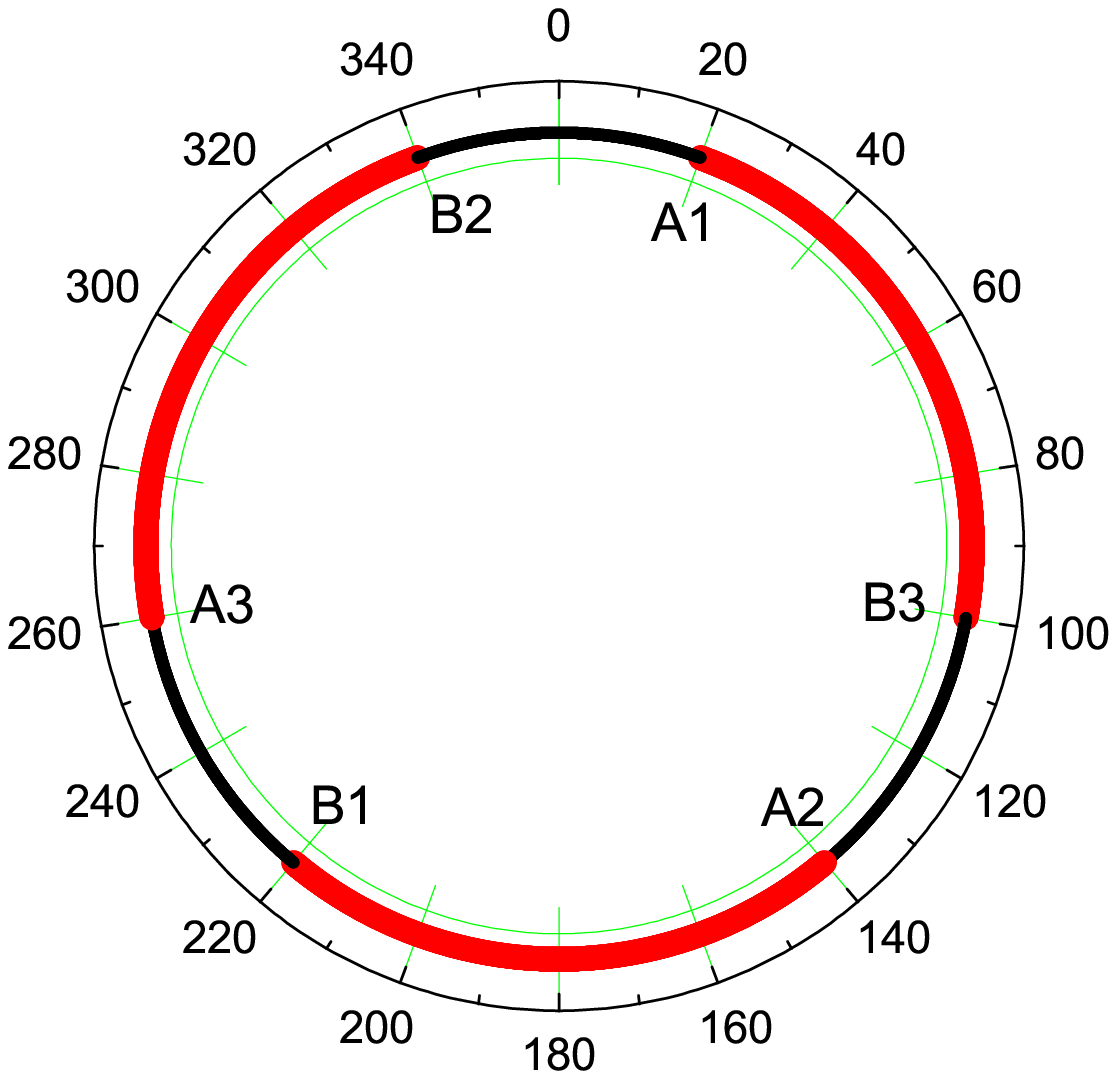}
  \caption{Schematic diagram of Plutinos' azimuthal distribution (the unit is degree), for the case of the resonance amplitudes $A\le 120^{\circ}$. This diagram shows three overlapping Plutino-arcs: A$1$-B$1$, A$2$-B$2$, and A$3$-B$3$. Each arc represents the combination of two adjacent red curves with a black curve between them.}
  \label{fig:plu}
\end{figure}

To establish the initial conditions for the Plutinos, we uniformly select the resonant angles $\sigma$ from $180^{\circ}-A_{max}$ to $180^{\circ}+A_{max}$, where $A_{max}$ represents their maximum resonant amplitude. And, the longitudes of perihelia $\varpi$ are randomly chosen between 0 and $360^{\circ}$. Then, given Neptune's mean longitude $\lambda_N$ at the beginning of the calculation, we can determine the initial mean longitudes $\lambda$ of the Plutinos using Eq. (\ref{longitude}). It is essential to note that when unfolding the phase space of the 2:3 resonance, three resonant islands emerge, corresponding to $\sigma\in$ [0, $360^{\circ}$], [$-360^{\circ}$, 0] and [$360^{\circ}$, $720^{\circ}$], respectively \citep{li2023study}. Therefore, besides the commonly considered range of [0, $360^{\circ}$], $\sigma$ is allowed to vary within the other two ranges by associating the angle $\sigma=\sigma\pm360^{\circ}$. Although such displacements in $\sigma$ preserve the resonant behaviours of Plutinos, the measurements of their longitude positions relative to Neptune, i.e. $\Delta\lambda$, would significantly change. For example, with $A_{max}=120^{\circ}$ and an initial $\lambda_N=0$, the initial values of $\lambda$ are distributed within three intervals of $[20^{\circ}, 220^{\circ}]$, $[140^{\circ}, 340^{\circ}]$, and $[260^{\circ}, 360^{\circ}]\bigcup[0, 100^{\circ}]$. Figure \ref{fig:plu} sketches the resulting azimuthal distribution of Plutinos, where the three arcs (A$1$-B$1$, A$2$-B$2$, and A$3$-B$3$) corresponds to the three respective $\lambda$ intervals. These arcs overlap at the regions indicated by the red thicker curves. So obviously, the azimuthal distribution of Plutinos is better described using the three-arc model rather than the ring model.

For Plutinos with zero-inclination orbits, besides $\lambda$ and $\varpi$, two additional orbital elements, the semimajor axis $a$ and the eccentricity $e$, need to be set. Initially, all Plutinos are placed at $a=39.4$ AU, corresponding to the nominal location of Neptune's 2:3 resonance, with varying initial values of $e$. Bearing in mind that the initial $\lambda$ is derived from the resonant angle $\sigma$ via Eq. (\ref{longitude}), and given that $\sigma\in [180^{\circ}-A_{max}, 180^{\circ}+A_{max}]$, these Plutinos form three distinct arcs, as illustrated in Fig. \ref{fig:plu}. The azimuthal lengths of these arcs are determined by $A_{max}$, while their curvatures (equivalent to heliocentric distances of Plutinos) are influenced by the value of $e$. Thus, the maximum resonant amplitude $A_{max}$ and the eccentricity $e$ of Plutinos serve as the two adjustable parameters of our three-arc model. We examine the orbits of observed Plutinos from the MPC and then choose representative values for these two parameters within the ranges of $A_{max}\le120^{\circ}$ and $e=0.05-0.35$. In addition, our earlier work \citep{li2014study} shows that an upper limit of $A=120^{\circ}$ is crucial for the stability of Plutinos over the age of the solar system. Regarding the the lower limit of $e=0.05$, since the strength of the 2:3 resonance is proportional to $e$, Plutinos with $e<0.05$ are generally too weak to sustain their resonant behaviours. In fact, $A_{max}$ and $e$ can respectively introduce asymmetries in the azimuthal and radial distributions of Plutinos, resulting in different magnitudes of changes in the Sun-Neptune distance. Interestingly, the combined effects of these two orbital parameters can possibly weaken the total perturbation of Plutinos, as we will show below.

\begin{figure*}
\includegraphics[width=\columnwidth]{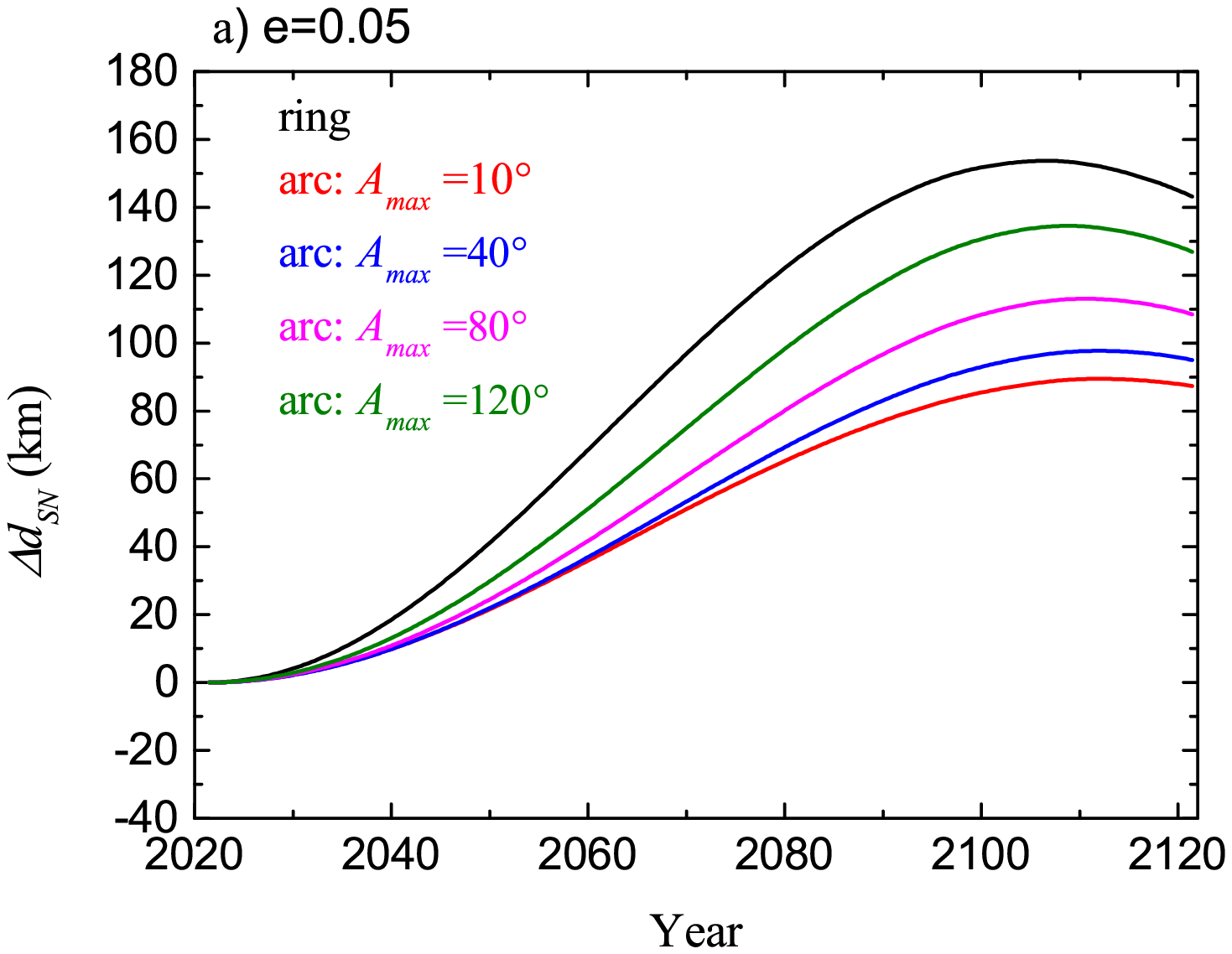}
\includegraphics[width=\columnwidth]{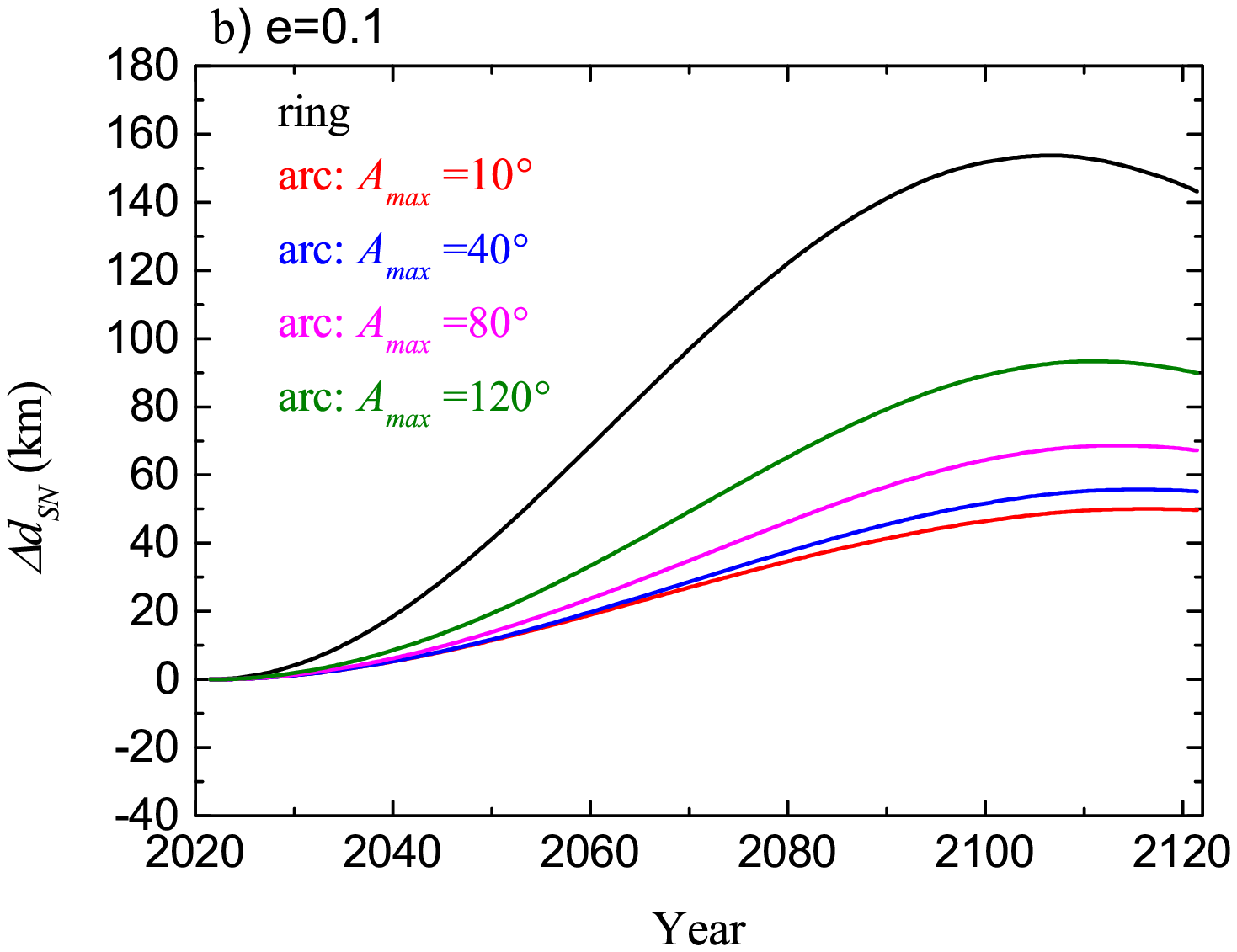}
\includegraphics[width=\columnwidth]{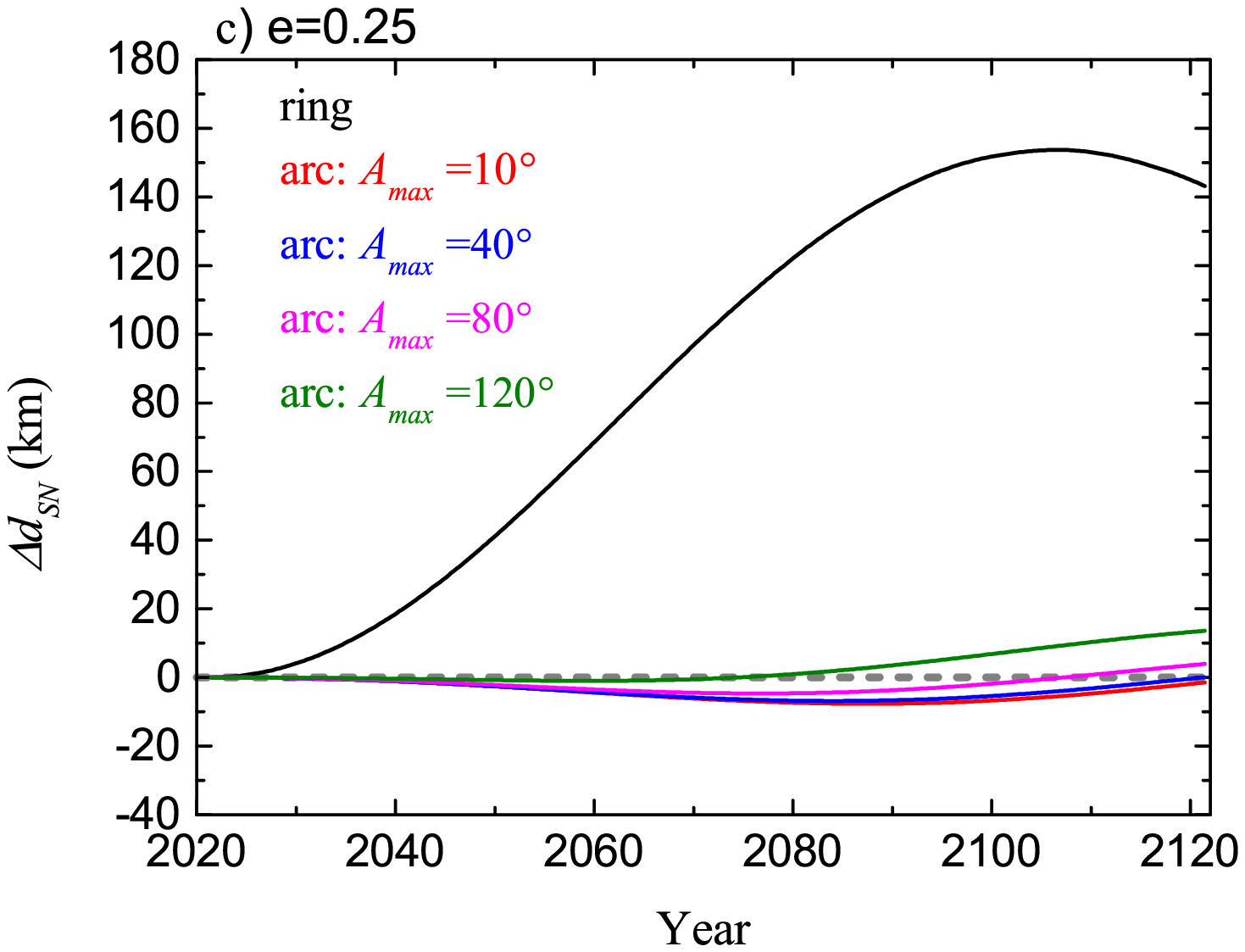}
\includegraphics[width=\columnwidth]{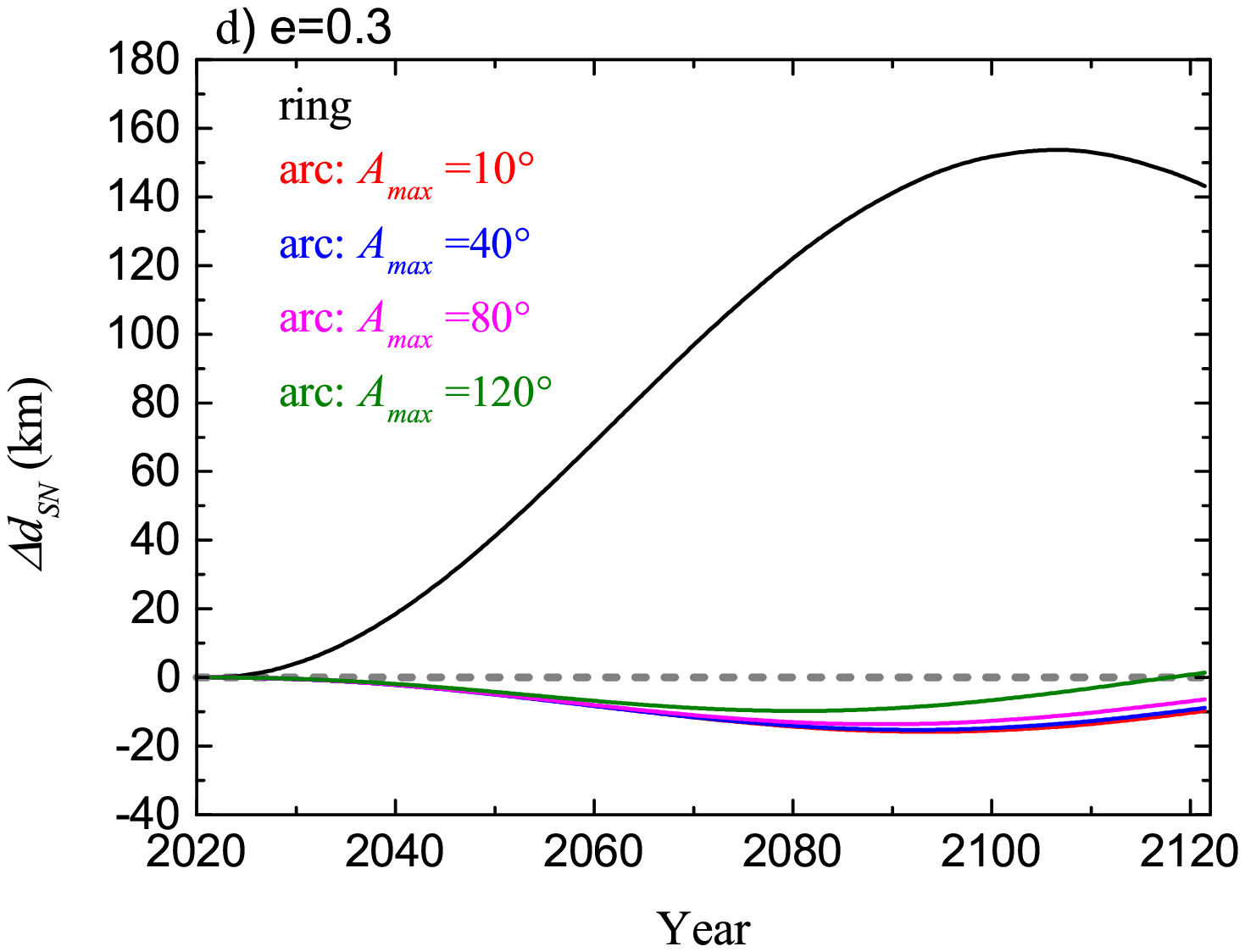}
\caption{Perturbation on the Sun-Neptune distance induced by the three arcs, within which Plutinos possess resonant amplitudes of $A \le A_{max}$ and varying eccentricities, as: (a) $e=0.05$, (b) $e=0.1$, (c) $e=0.25$, and (d) $e=0.3$. The black curve represents the same reference case of a continuous ring fixed at 39.4 AU, while the colourful curves refer to the arcs with $A_{max}=10^{\circ}$ (red), $40^{\circ}$ (blue), $80^{\circ}$ (magenta), and $120^{\circ}$ (olive). In the two lower panels, a horizontal dashed line is plotted at $\Delta d_{SN}=0$, which corresponds to the unperturbed solar system model.}
\label{fig:3}
\end{figure*}

To ensure an adequate number of Plutinos in each discrete arc to effectively replace the continuous arc, we consider the case of $A_{max}=120^{\circ}$. In this scenario, the azimuthal length of each arc reaches its maximum of $200^{\circ}$, as depicted by the individual arcs A$i$-B$i$ ($i=1,2,3$) in Fig. \ref{fig:plu}. Let $l=200^{\circ}=10\pi/9$ and $r=39.4$ AU in Eqs. (\ref{number_ring}) and (\ref{number_arc}), we can obtain that a minimum number of $n_{arc}\approx3000$ Plutinos is needed for representing a single arc. Accordingly, in subsequent 100-yr calculations, we employ 9000 equal point masses for our three-arc model, regardless of $A_{max}$. The total mass of these point masses is still adopted to be $M_{plu}=0.01M_\oplus$, as we did for the ring model.

Before concluding this subsection, we pause to discuss the model's centre of mass. For the continuous ring model analytically described by Eq. (\ref{eq:ring}), the centre of mass can be fixed at the Sun \citep[e.g.][]{Kuchynka2010}. However, for the discrete arc model consisting of a number of point masses, especially for the eccentric one, it is not feasible to simply maintain its centre of mass in the same manner. Instead, the arcs' orbits are bound to the Sun by gravitation. A similar approach was previously used for the point-mass models in \citet{pitjeva2018mass} and \citet{di2020analysis}. As stated at the beginning of this section, the advantage of this approach is that the arcs can naturally revolve alongside Neptune, better mimicking the motion of observed Plutinos. We acknowledge that the centre of mass of our arc model could drift, here by a value of less than 1 km. Similarly, the center of mass of observed Plutinos would also drift. In Sect. \ref{sec:real} we will see that the arc model indeed effectively represents the perturbations of observed Plutinos.

\subsection{Perturbation difference between the ring and arc models}
\label{result} 

To compare the total perturbations of Plutinos modeled by the ring and arc models, we calculate the corresponding changes $\Delta d_{SN}$ in the Sun-Neptune distance. In the ring model, $\Delta d_{SN}$ is derived from the continuous ring described by Eq. (\ref{eq:ring}), where the radius $r$ is fixed at 39.4 AU. On the other hand, in the arc model, $\Delta d_{SN}$ is obtained from three discrete arcs that we constructed in Sect. \ref{subsec:arc}. These arcs are also located at 39.4 AU but have varying orbital parameters $A_{max}$ and $e$ to account for the specific distribution of Plutinos.

The temporal evolutions of $\Delta d_{SN}$ are plotted in Fig. \ref{fig:3}, covering a time span of 100 yr from 2021 to 2121. In each panel (a)-(d), we present the same $\Delta d_{SN}$ variation for the ring model (depicted by the black curve), which will be considered as the reference case. It is observed that the change in the Sun-Neptune distance induced by the Plutino-ring perturbation experiences a substantial increase shortly after the beginning in 2021, reaching a maximum value of $\Delta d_{SN}\sim154$ km around the year 2100. Subsequently, $\Delta d_{SN}$ slightly decreases to about 143 km by the end of our calculation at the epoch 2121.

Next, we analyze the three-arc model, beginning with the case of $e=0.05$, as shown in Fig. \ref{fig:3}(a). To better understand the trend in the variation of $\Delta d_{SN}$ resulting from changes in the azimuthal distribution of Plutinos, we selectively plot colourful curves for representative resonant amplitudes of $A \le A_{max}=10^{\circ}$ (red), $40^{\circ}$ (blue), $80^{\circ}$ (magenta), and $120^{\circ}$ (green). For each given $A_{max}$, the Plutinos have their resonant angles $\sigma$ confined to the range of $[180^{\circ}-A_{max}, 180^{\circ}-A_{max}]$. One can see that for Plutinos with $A \le A_{max}=10^{\circ}$, around the year 2100, $\Delta d_{SN}$ reaches its peak of 89 km, approximately half of the maximum $\Delta d_{SN}$ obtained in the ring model (indicated by the black curve). But as $A_{max}$ increases from $10^{\circ}$ to $120^{\circ}$, the corresponding curve (from red to green) approaches the black one associated with the ring model, indicating a diminishing asymmetry in the azimuthal distribution of Plutinos. This phenomenon can be attributed to the fact that the $\sigma$-range widens at larger $A_{max}$ and becomes closer to the complete interval of $\sigma=[0, 360^{\circ}]$ as in the ring model. Nevertheless, even when $A_{max}=120^{\circ}$, the discrepancy in $\Delta d_{SN}$ between the ring (black curve) and arc (green) models can still exceed 10\%.

Figure \ref{fig:3}(b) illustrates the outcomes for another representative case of $e=0.1$. We can see that the difference between the $\Delta d_{SN}$ values derived from the ring and arc models gradually diminishes as $A_{max}$ increases. This trend in $\Delta d_{SN}$ with increasing $A_{max}$ remains apparent and similar to the case of $e=0.05$ (see Fig. \ref{fig:3}(a)). It suggests that, at relatively small $e$, the asymmetry of the azimuthal distribution of Plutinos in their gravitational perturbations could be notable.

However, when the eccentricities of Plutinos are large, the asymmetry in their radial distribution becomes important and can significantly impact the Sun-Neptune distance. Figures \ref{fig:3}(c) and (d) show the results for the cases of $e=0.25$ and 0.3, respectively. A remarkable pattern emerges as the colourful curves representing different $A_{max}$ values cluster around $\Delta d_{SN}=0$. These high-eccentricity cases highlight our arc model in two key respects: (1) the perturbation caused by Plutinos becomes exceptionally weak due the combined effects of the azimuthal and radial distribution asymmetries of Plutinos, determined by $A_{max}$ and $e$, respectively. Over the entire 100-yr evolution, the resulting effect on the Sun-Neptune distance differs by only 10-20 km compared to the unperturbed solar system model, where $\Delta d_{SN}=0$. A plausible explanation for the weak perturbation induced by Plutinos with large eccentricities could stem from their resonant configuration. When Plutinos reach their perihelia, they tend to cluster in two segments in longitude, leading and trailing Neptune by $\sim90^{\circ}$, respectively. As their eccentricities increase, the distances between Neptune and the Plutinos would increase during conjunctions \citep{Chiang2002}. Therefore, the gravitational effects of more eccentric Plutinos on Neptune become weaker. (2) Regardless of $A_{max}$, the colourful curves deviate prominently from the black curve, exhibiting differences as large as about 170 km in $\Delta d_{SN}$. This significant deviation clearly demonstrates that the ring model is completely unsuitable for accurately representing the perturbations induced by the Plutinos.

\begin{figure}
  \includegraphics[width=\columnwidth]{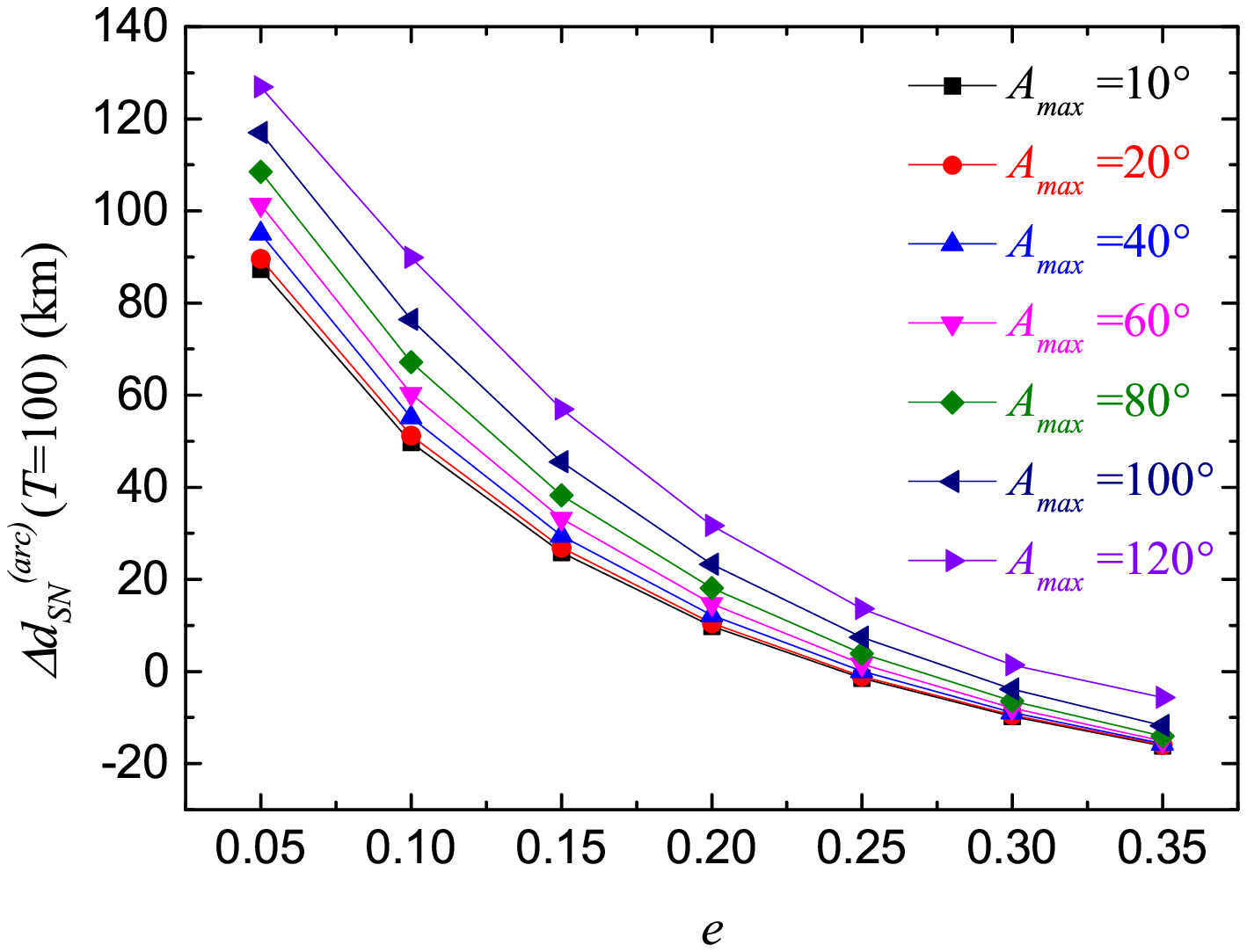}
  \caption{Dependence of $\Delta d_{SN}$ on the eccentricities $e$ of Plutinos for seven different $A_{max}$ values: $10^{\circ}$ (black), $20^{\circ}$ (red), $40^{\circ}$ (blue), $60^{\circ}$ (magenta), $80^{\circ}$ (green), $100^{\circ}$ (navy), and $120^{\circ}$ (purple). The values of $\Delta d_{SN}$ are generated using the three-arc model at the end of our 100-yr calculations (i.e. $T=100$).}
  \label{fig:5}
\end{figure}

Finally, we provide a concise description for the dependence of the perturbation of Plutinos on their parameters: the maximum resonant amplitude $A_{max}$, the eccentricity $e$, and the total mass $M_{plu}$. Figure \ref{fig:5} summarizes the changes in the Sun-Neptune distance induced by the three-arc model at the end of our 100-yr calculation, denoted as $\Delta d^{(arc)}_{SN}(T=100)$, for various values of $A_{max}$ and $e$. Through linear fitting of the data points corresponding to different $(e, A_{max})$ combinations in this figure, we establish the following measurement: 
\begin{equation}
    \Delta d^{(arc)}_{SN}(T=100)=\alpha e+\beta ~~~\mbox{(km)},
    \label{eq:deltadarc}
\end{equation}
where the coefficients $\alpha$ and $\beta$ are given by
\begin{equation}
\begin{split}
    \alpha&=-1.1 (A_{max} / 1^{\circ})-310.6, \\
    \beta&=0.4 (A_{max} / 1^{\circ})+78.7.
\end{split}
\end{equation}

\begin{figure}
\includegraphics[width=\columnwidth]{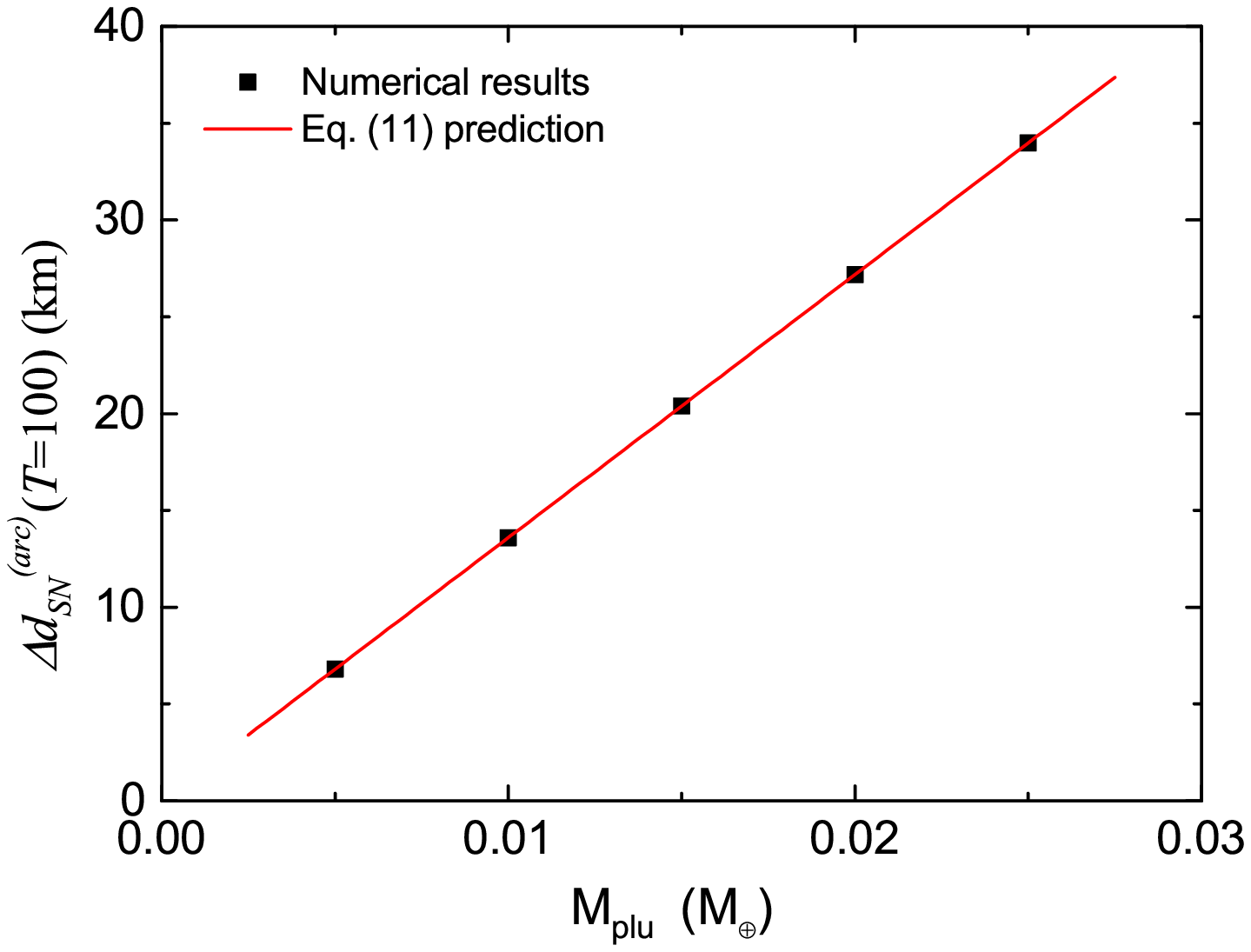}
\caption{Dependence of $\Delta d_{SN}$ on the total mass $M_{plu}$ of Plutinos, with $e=0.25$ and $A_{max}=120^{\circ}$ in the three-arc model. The squares indicate the results from numerical computations, and the red curve denotes the prediction from Eq. (\ref{eq:AeMt}). This figure illustrates good agreement between these two approaches. As in Fig. \ref{fig:5}, $\Delta d_{SN}$ refers to the epoch of $T=100$.}
  \label{fig:mass}
\end{figure}

Furthermore, the total mass $M_{plu}$ of Plutinos may be a crucial parameter in determining $\Delta d^{(arc)}_{SN}$. Although the precise value of $M_{plu}$ currently remains uncertain, the first-order approximation made in previous works \citep{Kuchynka2010, li2018constructing} suggests that the perturbations induced by point masses are nearly proportional to their masses. By incorporating the mass factor into Eq. (\ref{eq:deltadarc}), we can derive a comprehensive expression for the perturbation of Plutinos on the Sun-Neptune distance:
\begin{equation}
\Delta d^{(arc)}_{SN}(T=100)=\frac{M_{plu}}{0.01M_\oplus}(\alpha e+\beta) ~~~\mbox{(km)}.
\label{eq:AeMt}
\end{equation}
To validate Eq. (\ref{eq:AeMt}), we selected typical parameters of $e=0.25$ and $A_{max}=120^{\circ}$, and varied the total mass $M_{plu}$ in the range of 0.005-0.025 $M_\oplus$. First, we numerically recalculate the change in the Sun-Neptune distance, i.e. $\Delta d_{SN}$, using the three-arc model with different $M_{plu}$ values. The resulting $\Delta d_{SN}$ at the end of the 100-yr integration is indicated by the squares in Fig. \ref{fig:mass}. Then, for various total masses $M_{plu}$, we predict $\Delta d^{(arc)}_{SN}$ at $T=100$ via Eq. (\ref{eq:AeMt}). The predictions are depicted by the red curve. 
Notably, good agreement between the numerical results and our predictions is evident, as the absolute error is less than $10^{-4}$ km. This error is several orders of magnitude smaller than the obtained $\Delta d_{SN}$ value or the difference with the continuous versus discrete ring modularization proposed in Sect. \ref{subsec:ring}. Accordingly, the nearly linear dependence of $\Delta d_{SN}$ on $M_{plu}$, as described in Eq. (\ref{eq:AeMt}), can be nicely supported.

At this point, we propose that the parameter $M_{Plu}$ can be decoupled from the other parameters (i.e. $e, A_{max}$) in our arc model. This implies that when the total mass $M_{Plu}$ of Plutinos is refined in future surveys, one can easily estimate their influence on the Sun-Neptune distance using a linear relation in $M_{Plu}$ instead of conducting repetitive numerical computations.

\section{Comparison with observed point-mass Plutinos}
\label{sec:real}
As we proposed, the impact of the eccentric Plutino orbits has to be modeled by a non-uniformly distributed mass, i.e. the eccentric arcs, over a circular ring. Naturally, we need to verify that the three-arc model is sufficient to represent the perturbations induced by the individual contributions of observed Plutinos on eccentric orbits. To do so, we will compare the positions of the planets perturbed by the three-arc model and those perturbed by the point-mass eccentric Plutinos from observations.

As of May 2024, over 3200 KBOs have been registered in the MPC database\footnote{https://minorplanetcenter.net/iau/lists/TNOs.html}. For the sake of identifying the Plutinos, we first consider the KBOs with semi-major axes of $39\le a \le 40$ AU. By this criterion, 519 KBOs are selected to be Plutino candidates. Then the orbits of these candidates are numerically integrated in a solar system model consisting of the Sun and four Jovian planets. Finally, an object is regarded as a Plutino if its resonant angle $\sigma$ librates, i.e. the resonant amplitude of $A<180^\circ$. The total duration of the integration is chosen to be 1 Myr. We note that the typical libration period of the 2:3 resonant angle $\sigma$ is about 20,000 yr; for example, Pluto has a resonant period of 19,670 yr \citep{1965AJCohen,murray_dermott_2000}. Therefore, the integration time of 1 Myr is approximately equal to 50 resonant periods, which is long enough to identify the Plutinos that can stably exhibit the libration behaviour of $\sigma$.

\begin{figure}
\includegraphics[width=\columnwidth]{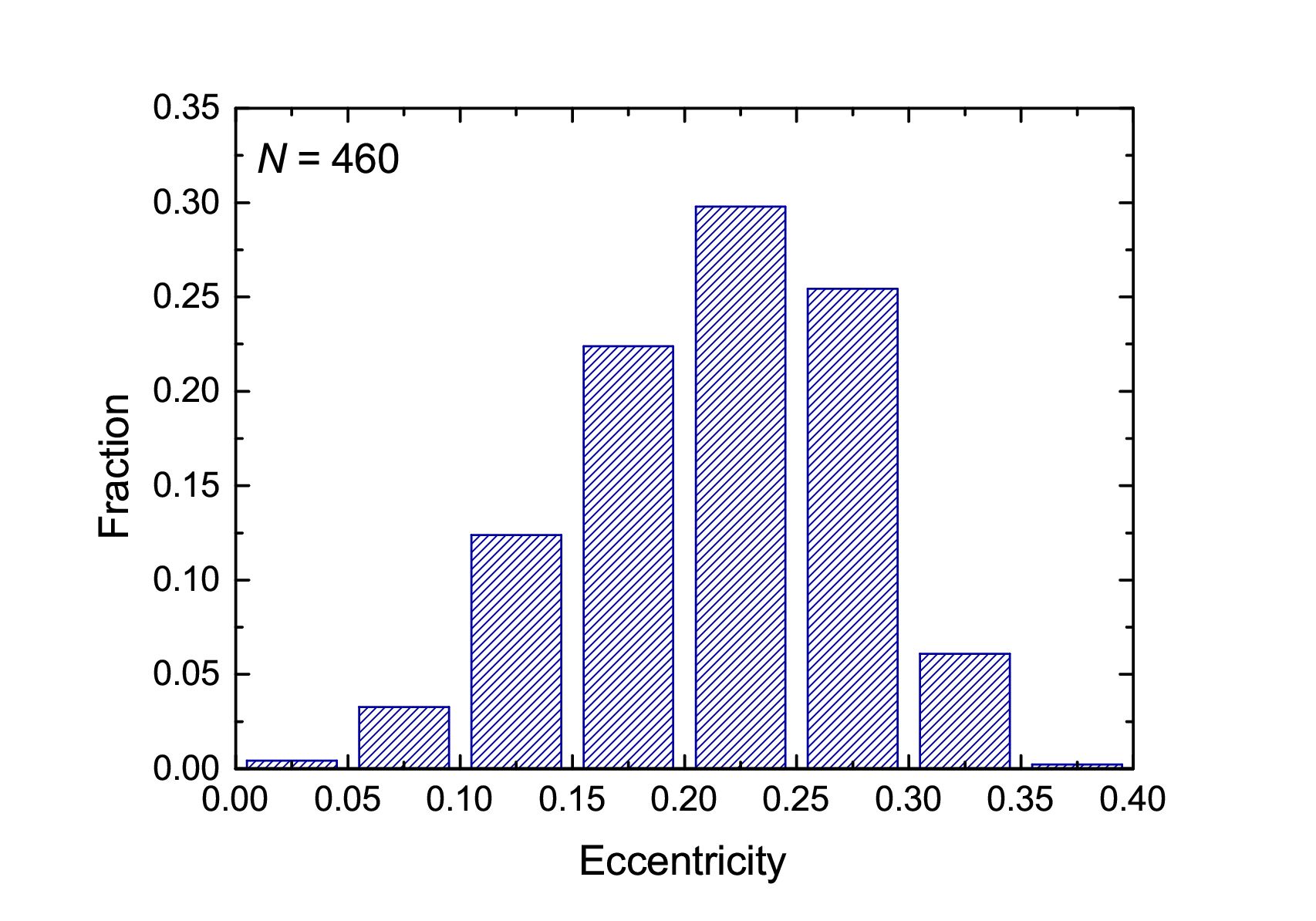}
\caption{The eccentricity distribution of the currently observed Plutinos (as of May 2024).}
  \label{fig:6_1}
\end{figure}

Eventually, 460 Plutinos have been identified, and their eccentricity distribution is shown in Fig. \ref{fig:6_1}. Considering that the arc model shows a greater difference from the ring model at high eccentricity, we will focus on the eccentricity range of $e>0.2$ below. There are 278 Plutinos that belong to this $e$ range, accounting for more than 60\% of the entire Plutino population. 
It is evident that the number of currently observed Plutinos is much smaller than the theoretically estimated population, which is over 9000 for diameters $D>100$ km, as mentioned in Sect. \ref{sec:intro}. This number deficit is because surveys for KBOs are usually limited in relatively small areas ($\sim 100$ square degrees) due to their faint magnitudes ($m_R\sim23$ with $D\sim100$ km) \citep{bannister2018ossos,Bernardinelli_2020,2022DPS....5450101T}. 
Therefore, the observational incompleteness of the available observed Plutinos can lead to a biased spatial distribution. 
The spatial distribution of Plutinos is crucial as it significantly affects the estimation of their total perturbation. Furthermore, for the Plutinos with $e>0.2$, they are more likely to be observed when they reach their perihelia and are at their brightest. However, given the orbital period of the Plutino population, which is about 250 yr long, those objects that are near their aphelia, i.e. at relatively larger heliocentric distance, would be more difficult to observe currently and in the near future.

In order to offset the observational bias, we have constructed two methods to simulate the realistic spatial distribution of the overall Plutinos:

(1) Dispersion in space (DISpa): Since the number of Plutinos currently observed is much less than theoretically predicted, we enlarge the sample size following the idea of the Monte Carlo method \citep{li2019calibration,li2020Planet9}. It is obvious that the spatial dispersion of the Plutino samples strongly depends on the three angles of their orbital elements: longitude of ascending node ($\Omega$), argument of perihelion ($\omega$), and mean anomaly ($M$). Thus, for each of the 278 Plutinos with $e>0.2$, we generate 10 synthetic samples with the same $a$, $e$, and $i$, as well as with the same resonant angle $\sigma$, but their $\Omega$ and $\omega$ are randomly selected between $0^{\circ}$ and $360^{\circ}$. The mean anomaly $M$ of a synthetic sample is determined according to Eq. (\ref{eq:resonantargument}). 
In this method, we produced a more `complete'  Plutino population, consisting of 2780 synthetic Plutino samples, referred to as DISpa samples.

(2) Dispersion in time (DITim): Regarding the current 278 Plutinos with $e>0.2$, we numerically integrated their trajectories for $10^7$ yr, under the gravitational perturbations of four Jovian planets. Since the $10^7$-yr evolution time is much longer than the secular procession period of the orbits of Plutinos \citep{li2014study}, this method can naturally disperse the observed Plutinos within the six-dimensional orbital distribution, even with a sample size as small as 278. These samples will be denoted by the DITim samples.

It is worth mentioning that the inclinations of the 278 considered Plutinos are taken to be the observed values, meaning that both the DISpa and DITim samples generated from this population follow an inclination distribution similar to the observations. This may further allow us to test the validity of the theoretical arc model, which is co-planar in principle.

\begin{figure*}
\includegraphics[width=\columnwidth]{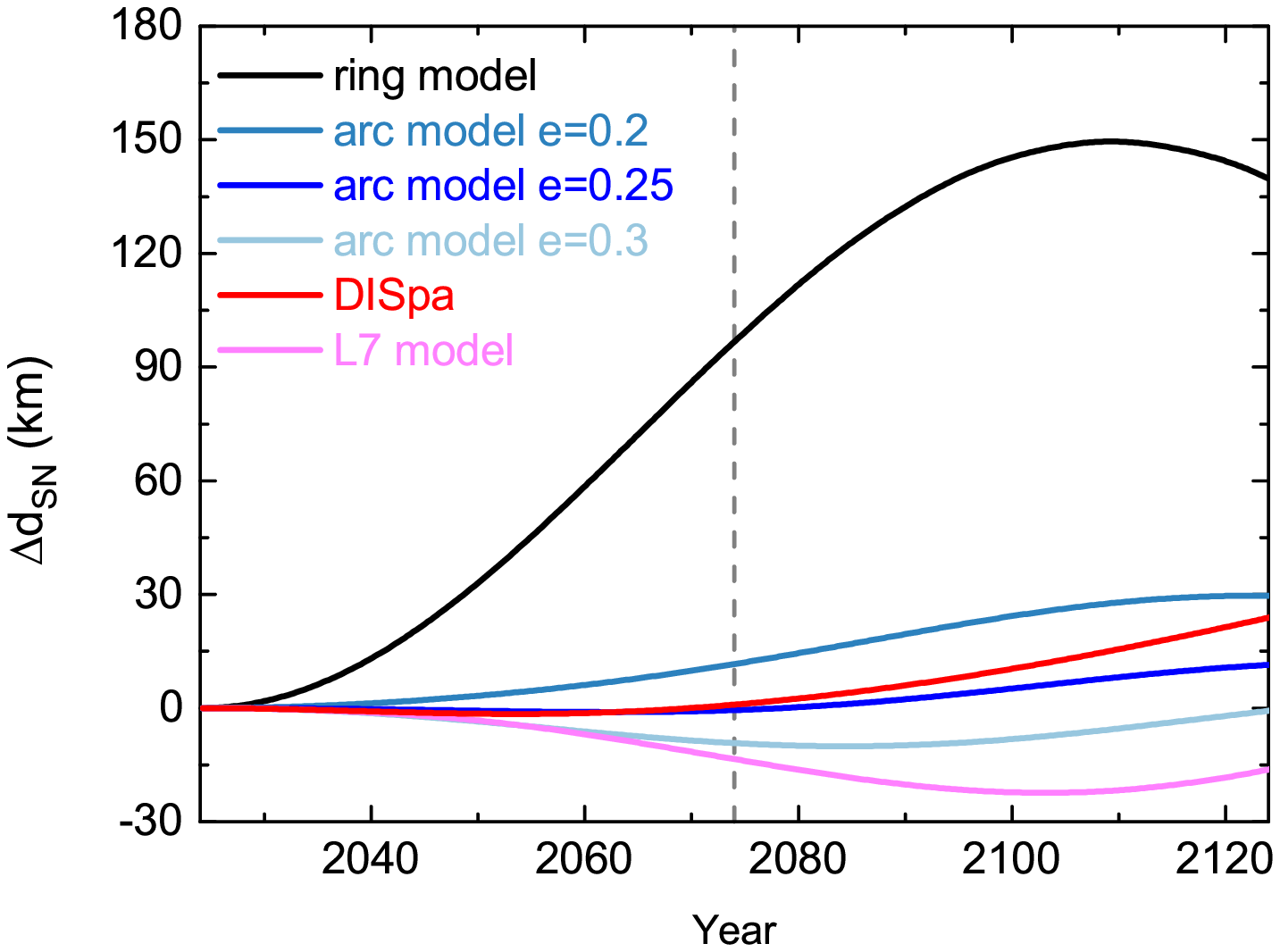}
\includegraphics[width=\columnwidth]{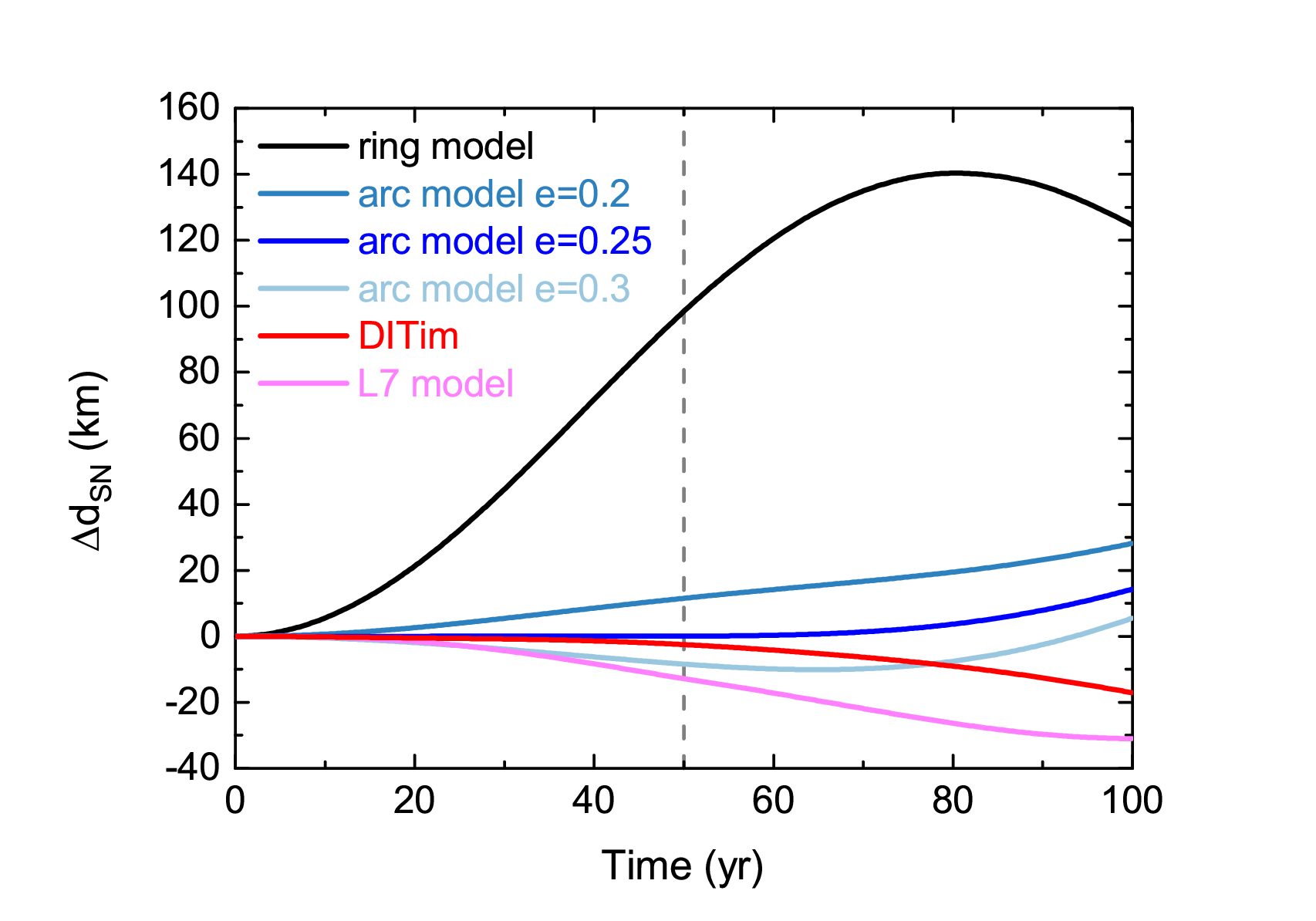}
\caption{Perturbations on the Sun-Neptune distance induced by the arc models with eccentricities of $e=0.2$ (steel blue), $e=0.25$ (regular blue), $e=0.3$ (light blue), the ring model (black), and the point-mass Plutinos with $e>0.2$ (red). The spatial distribution of the point-mass Plutinos is corrected by the DISpa (left panel) or DITim (right panel) method. In addition, a case using the point-mass Plutinos selected from the L7 synthetic model is also presented (light magenta curve). Note that the x-coordinates differ between the left and right panels due to the distinct initial epochs of the modeled planetary systems, in the year 2024 and $10^7$-yr into the future, respectively. Consequently, the arc models do not give the exact same results on these two panels. For reference, the vertical dashed line is plotted to mark the 50 yr of evolution.}
\label{fig:SNPlu}
\end{figure*}

Subsequently, we calculated the perturbation on the Sun-Neptune distance induced by the point-mass Plutinos from the two methods described above.The results are shown by the red curves in Fig. \ref{fig:SNPlu}, with the left and right panels corresponding to the DISpa and DITim samples, respectively.
This figure also displays the Sun-Neptune distance difference $\Delta d_{SN}$ induced by the ring model (black curve) and the arc models with eccentricities $e=0.2$, 0.25 and 0.3 (indicated by different shades of blue). It can be seen that, in general, the time evolution of $\Delta d_{SN}$ for the point-mass Plutinos deviates significantly from that of the ring model, but it is much closer to that of the arc models.
Specifically, the arc model with $e=0.25$ nicely mimics the point-mass Plutinos, as the corresponding regular blue and black curves have a difference in $\Delta d_{SN}$ of less than 2 km for the first 50 yr. This could be understood by acknowledging that, as shown in Fig. \ref{fig:6_1}, the eccentricity $e=0.25$ corresponds approximately to the average eccentricity of the considered Plutinos with $e>0.2$.

However, after 50 yr of evolution, the difference in $\Delta d_{SN}$ between the cases of the point-mass Plutinos and the arc model with $e=0.25$ increases, reaching about 20 km and 40 km for the DISpa and DITim samples, respectively. This noticeable discrepancy may be due to the fact that observational bias has not been completely eliminated. Nevertheless, we find that the difference in $\Delta d_{SN}$ between the cases of the point-mass Plutinos and the ring model is much larger, exceeding 140 km. We note that observations of the farthest planet, Neptune, primarily rely on astronomical observations. Due to limitations imposed by Earth's atmosphere and the accuracy of star catalogs, the orbital accuracy of Neptune is currently at the level of a few thousand kilometers \citep{folkner2014planetary,pitjeva2018mass,Stone_2001}. While the 140 km difference in Neptune's position that we derived may seem small compared to current observations, it is noteworthy because a similar magnitude of orbital variation, $\Delta d_{SN}$, will be introduced once the ring model developed in previous works incorporates the planetary ephemerides. Furthermore, our arc model will become increasingly valuable as future in situ observations lead to significant improvements in the positional accuracy of Neptune.

Having already taken into account the comparison of the 3-arc model to the observed Plutinos, represented by the DISpa and DITim samples. As a supplement, a new experiment has been conducted, letting the observed Plutinos be represented by the population of 2:3 resonant KBOs from the L7 synthetic model \citep{L7Kavelaars2009,L7Petit2011,L7Gladman2012}. The L7 synthetic model, based on the Canada France Ecliptic Plane Survey (CFEPS), gives a distribution of KBOs extrapolated down to small sizes, with $H_g$ magnitude of 8.5. Consistent with the DISpa and DITim samples, here we consider the L7 model samples with $e>0.2$, resulting in a total number of about 1500 objects. These objects were then used to calculate the perturbation on the Sun-Neptune distance, and the results are shown in Fig. \ref{fig:SNPlu} (see the light magenta curves). We note that, in relation to the right panel, we consider the L7 model samples after a $10^7$-yr evolution. In general, the $\Delta d_{SN}$-curve resulting from the L7 model is similar to that from the DISpa (or DITim) samples, tending to align more closely with the 3-arc model while diverging significantly from the ring model. One may notice a slight discrepancy: at an earlier epoch of about 30 yr, the time evolution of $\Delta d_{SN}$ for the L7 model starts to appear different from that for the 3-arc model. 
This discrepancy could be accounted for by the L7 synthetic model being derived solely from the CFEPS samples, which contain only 24 observed Plutinos. 
In fact, the CFEPS project website\footnote{https://www.cfeps.net/?page\underline{~}id=105} states that the L7 model may not represent the intrinsic orbit distribution of the KBOs, and direct comparisons with other KBO models should be avoided. Nevertheless, it is worth remarking that even within this 30-yr window, when referring to the L7 model, the ring model induces a maximum difference in $\Delta d_{SN}$ of over 40 km, which is ten times larger than the difference of about 4 km observed with the 3-arc model. This suggests that the 3-arc model is still a better representation of the observed Plutinos compared to the ring model.

Throughout the above analysis, three points need to be noted. First, to maintain consistency with the ring and arc models considered in Sect. \ref{sec:perturbation}, we assume the total mass of the eccentric Plutino samples to be equivalent to that of the entire Plutino population, i.e. $0.01M_\oplus$. In reality, however, the former should be less massive than the latter. Since $\Delta d_{SN}$ nearly linearly depends on the total mass of Plutinos, a smaller total mass could further reduce the absolute discrepancy between the point-mass Plutinos and our arc model. Secondly, in the construction of the DISpa samples, the initial conditions of the system are adjusted to be at the epoch of the year 2024, which is the same as the epoch of the KBOs registered in the MPC database. Therefore, the initial epoch in Fig. \ref{fig:SNPlu} is 3 yr later than that in Fig. \ref{fig:3}. Thirdly, the x-coordinates are different in the left and right panels in Fig. \ref{fig:SNPlu}. This is due to the distinct initial epochs of the modelled planetary systems. As mentioned above, in the left panel, the initial epoch of the DISpa samples--in the year 2024--represents the current time of the observed Plutinos. Meanwhile, in the right panel, the DISpa samples refer to the observed Plutinos after a $10^7$-yr evolution, meaning the initial epoch corresponds to $10^7$-yr into the future. Consequently, since the arc models in these two panels refer to planetary systems at different epochs, the resulting profiles of the $\Delta d_{SN}$ curves are not the same. We believe this epoch difference can further support the comparison between the point-mass Plutinos, the arc model, and the ring model. A similar rationale also applies to Fig. \ref{fig:SSPlu}.

In a brief summary, we believe that our three-arc model has a significant advantage in representing the total perturbation induced by observed Plutinos, especially for those on the eccentric orbits, which comprise a majority of the Plutino population.

Having constructed the Plutino models, we apply them to evaluate the impact of the Plutinos on Neptune's longitude and latitude. This could be useful because observational constraints for Neptune, the farthest planet, are often given in these two parameters. By making simple coordinate transformations on our previous results from the ring model, the arc model, and the DISpa Plutino samples, in Table \ref{table:2}, we provide the maximum changes in Neptune's longitude ($\Delta \lambda_N$) and latitude ($\Delta \phi_N$). For reference, the corresponding maximum changes in the Sun-Neptune distance ($\Delta d_{SN}$) are also included. The numbers in Table \ref{table:2} are measured within the first 50 yr of evolution, rather than a century. This shorter timescale is chosen because, as shown in Fig. \ref{fig:SNPlu}, the arc model better mimics the observed Plutinos during the first 50 yr; afterwards, the difference in $\Delta d_{SN}$ may become more evident. One may notice that, in comparison with the observational accuracy of geocentric longitudes and latitudes, the numbers given in Table \ref{table:2} appear pretty small for all the modeling. However, as we argued above, this does not undermine the applicability of our arc model relative to the ring model, as the latter could yield much larger differences in Neptune's position when compared to the observed Plutinos. Additionally, the consideration of Saturn's positions will further validate the arc model, as discussed in the following subsection.

\begin{table}
    \centering
    \begin{tabular}{l|c|c|c|c}
        \hline
       & $\Delta \lambda_N$ (mas) & $\Delta \phi_N$ (mas) & $\Delta d_{SN}$ (km) \\
           \hline
       ring model             & -6.70 &  0.55  &  98.17    \\
       arc model~($e=0.25$)   &  0.13 &  0.19  & -0.47    \\
       DISpa Plutino samples  &  0.56 & -0.03  & -1.04 \\
       L7 synthetic model & 0.65 & 0.13 & -13.5 \\
            \hline
    \end{tabular}
    \caption{The maximum changes in Neptune's longitude ($\Delta \lambda_N$), latitude ($\Delta \phi_N$), and the Sun-Neptune distance ($\Delta d_{SN}$) induced by different Plutino models during the 50 yr of evolution.}
    \label{table:2}
\end{table}

\subsection{Perturbation on Saturn positions}

\begin{figure*}
\centering
    \includegraphics[width=\columnwidth]{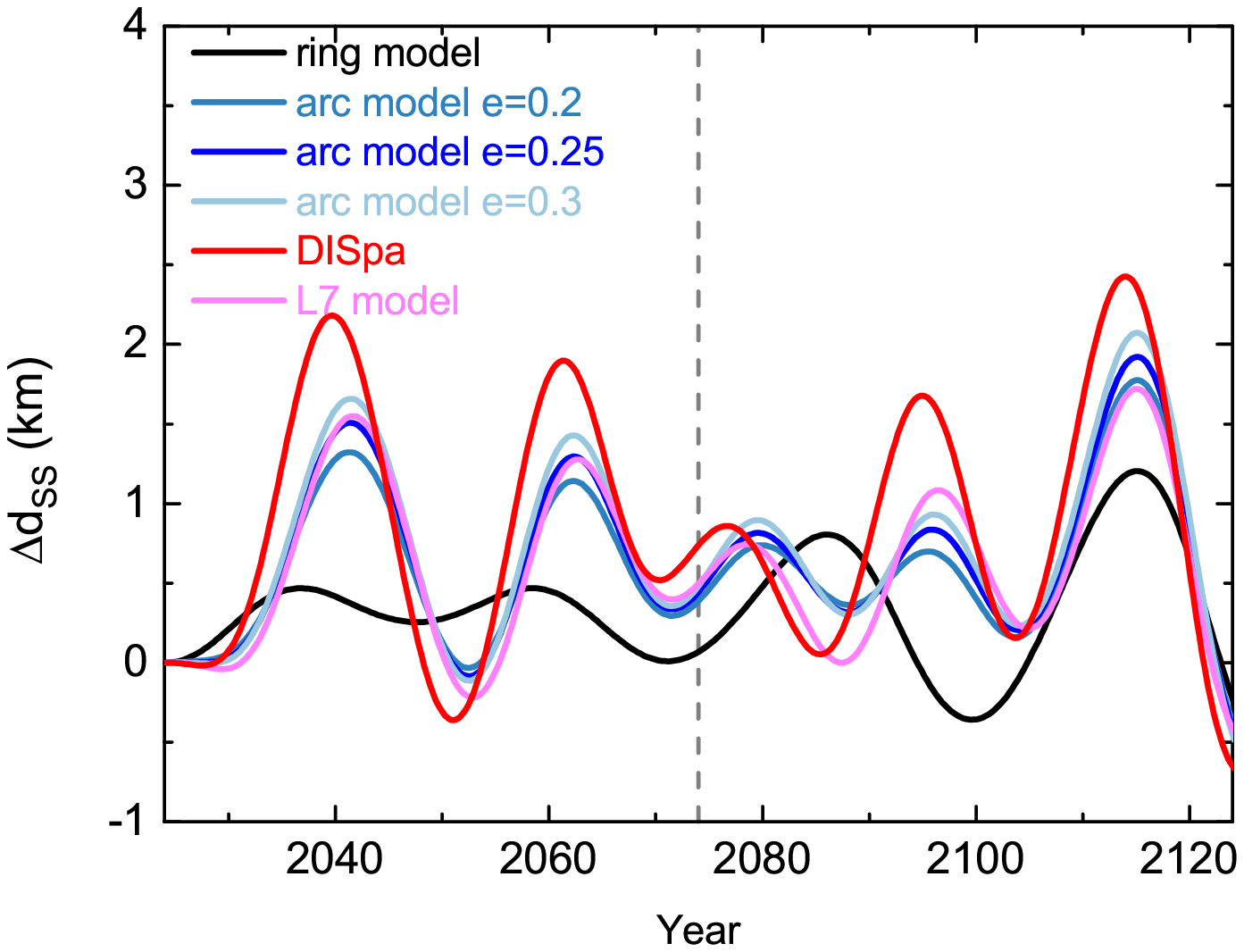}
    \includegraphics[width=\columnwidth]{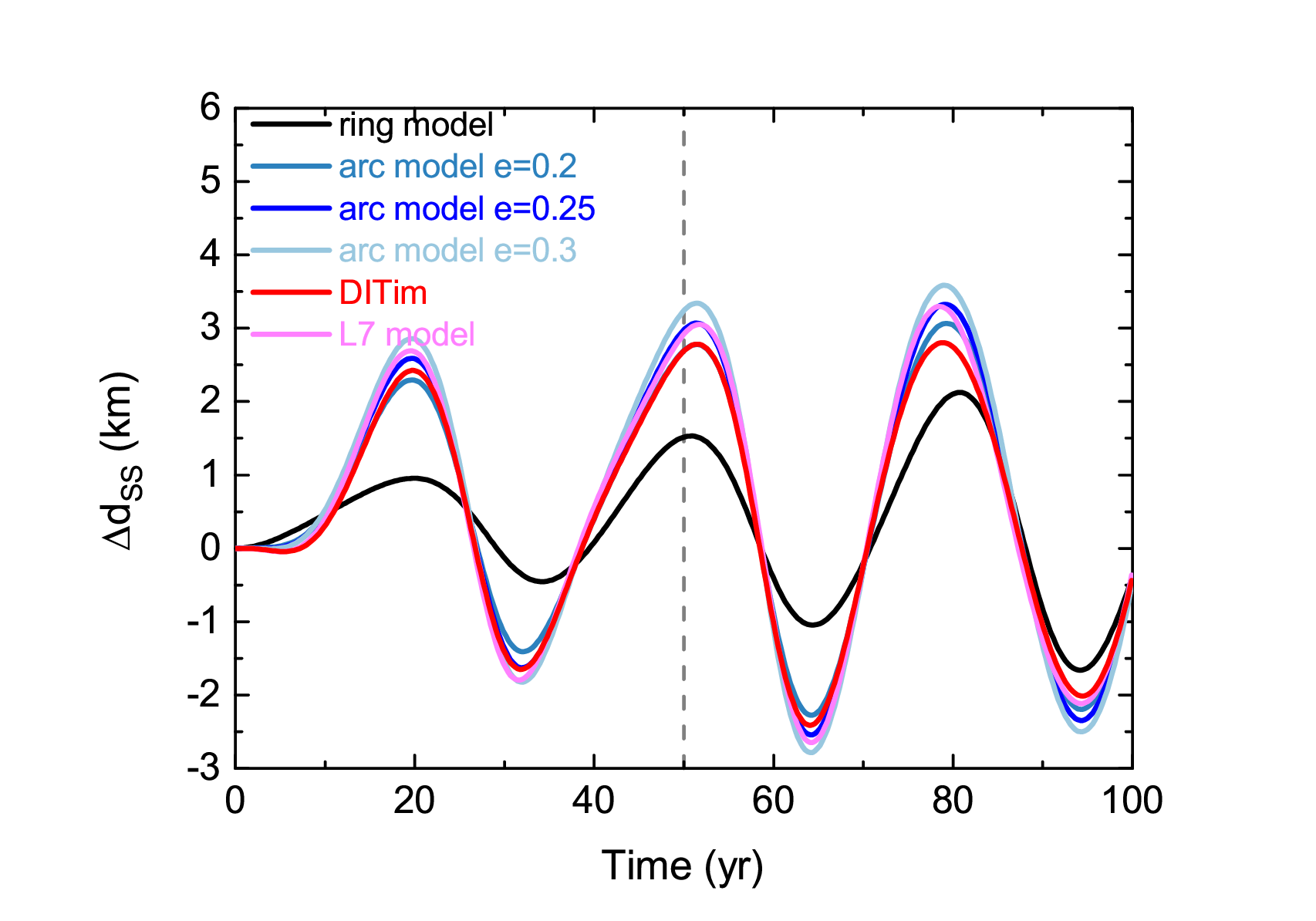}
\caption{Similar to Fig. \ref{fig:SNPlu} but for the change in the Sun-Saturn distance ($\Delta d_{SS}$).}
\label{fig:SSPlu}
\end{figure*}

\begin{figure*}
\centering
    \includegraphics[width=\columnwidth]{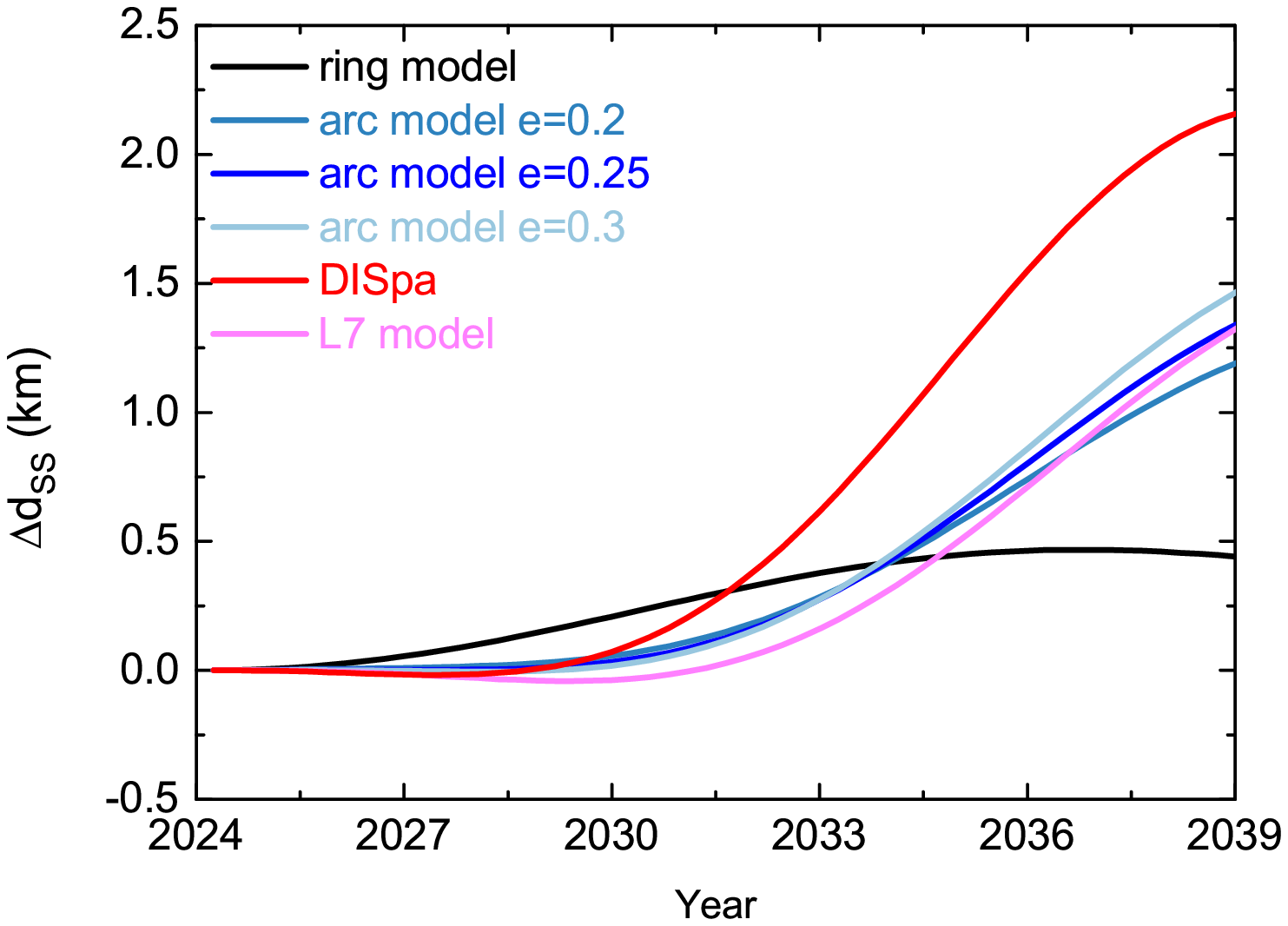}
    \includegraphics[width=\columnwidth]{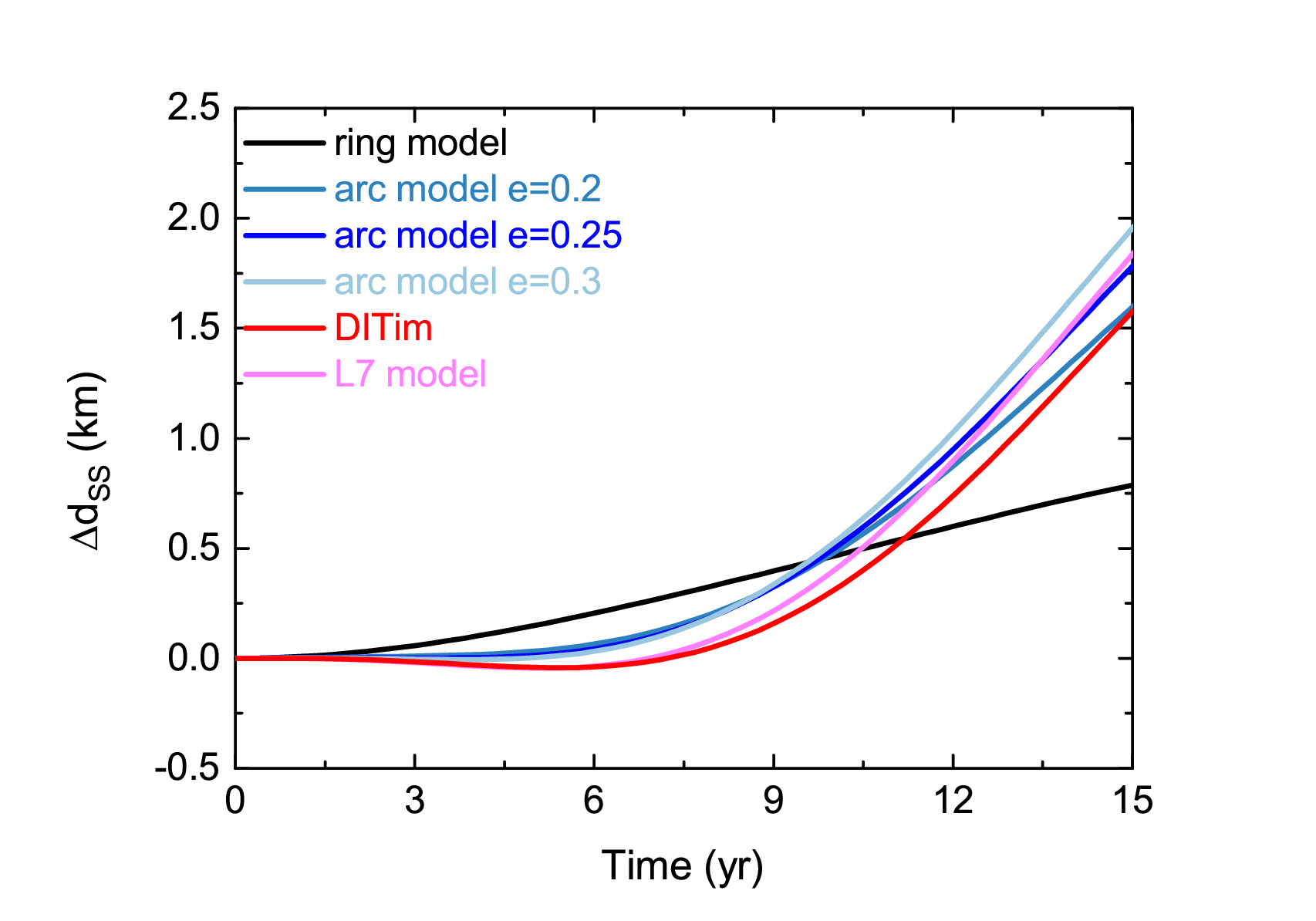}
\caption{A zoom-in view of Fig. \ref{fig:SSPlu}, showing the partial evolution of $\Delta d_{SS}$ within the first 15 yr.}
\label{fig:SSPlu15yr}
\end{figure*}

Compared to observations of Neptune, which rely mainly on astronomical observations, the accuracy of Saturn positions has been significantly enhanced by in situ observations. Based on a series of astrometric very long baseline interferometry (VLBI) observations of the \textit{Cassini} spacecraft, \citet{Jones2011} pointed out that the positional accuracy for Saturn is about 2 km. 
Later, based on \textit{Cassini} tracking data over the period between 2004 and 2017, the accuracy of the geocentric positions of Saturn was considerably improved to about 0.025 km \citep{Park_2021,fienga2021}. Therefore, present high-accuracy observations of Saturn positions may further verify the applicability of our three-arc model.

Results are shown in Fig. \ref{fig:SSPlu}, comparing the point-mass Plutinos, the arc model, and the ring model as in Fig. \ref{fig:SNPlu}, but this time plotted for the induced changes in the Sun-Saturn distance, denoted by $\Delta d_{SS}$. Similar to Fig. \ref{fig:SNPlu}, this figure also shows that the time evolutions of $\Delta d_{SS}$ for the point-mass Plutinos (red curve) and the arc models (blue curves) generally match well over the 100-yr evolution. However, there are considerable differences when compared to the ring model (black curve), in both the exact values and the curves' profiles.

Specifically, as seen in the left panel of Fig. \ref{fig:SSPlu}, when we consider the perturbation of the DISpa Plutino samples as the standard, the maximum difference in $\Delta d_{SS}$ for the arc model with $e=0.25$ is only about 0.96 km, while the difference for the ring model is more than twice as large, at approximately 2.15 km. Comparable results are achieved when taking the perturbation of the DITim Plutino samples as the standard, as shown in the right panel of Fig. \ref{fig:SSPlu}. The maximum differences in $\Delta d_{SS}$ are 0.87 km and 1.80 km for the arc model and ring model, respectively, with the latter being over twice as large. We again note that the different profiles of the corresponding $\Delta d_{SS}$-curves (e.g. the red curves) between the two panels in Fig. \ref{fig:SSPlu} are also due to the different initial epochs of the DISpa and DITim samples, as illustrated for Fig. \ref{fig:SNPlu}. Nevertheless, the comparisons among various models would not be affected.

Furthermore, to compare with the VLBI and \textit{Cassini} observations, Fig. \ref{fig:SSPlu15yr} presents a zoom-in view of Fig. \ref{fig:SSPlu} on the evolution of $\Delta d_{SS}$ over time. The 15-yr period shown corresponds approximately to the interval of \textit{Cassini} observations, as we mentioned above. Notably, during this period, the Sun-Saturn distance change, $\Delta d_{SS}$, has already reached a local maximum once. When comparing with the DISpa samples, as shown in the left panel of Fig. \ref{fig:SSPlu15yr}, the maximum difference in $\Delta d_{SS}$ induced by the arc model with $e=0.25$ is about 0.82 km, which is less than half of the 1.69 km difference produced by the ring model. And, the arc model exhibits even greater accuracy when compared with the DITim samples (right panel of Fig. \ref{fig:SSPlu15yr}), yielding a maximum $\Delta d_{SS}$-difference of only about 0.20 km. We additionally note that, although the L7 synthetic model was previously advised against direct comparison with other KBO models, its perturbation on Saturn's positions--indicated by the light magenta curve in Fig. \ref{fig:SSPlu15yr}--also seems to match the arc model (blue curves) quite well, with maximum differences in $\Delta d_{SS}$ ranging from 0.2 to 0.3 km. In contrast, the $\Delta d_{SS}$ difference between the L7 synthetic model and ring model consistently exceeds 1 km. Thus, the L7 synthetic model may further emphasize the significance of incorporating the arc model into planetary ephemerides instead of the ring model.

As our previous consideration of the changes in Neptune's longitude and latitude, Table \ref{table:3} presents similar results for Saturn, covering both the 50 yr and 15 yr evolutionary timescales. Over 50 yr of evolution, when compared to the point-mass Plutinos represented by either the DISpa samples or the L7 synthetic model, the arc model provides a noticeably better fit than the ring model for Saturn's longitude ($\Delta \lambda_S$), latitude ($\Delta \phi_S$), and the Sun-Saturn distance ($\Delta d_{SS}$). While for the 15-yr evolution, due to the shorter time span, the differences in $\Delta \lambda_S$ and $\Delta \phi_S$ are not significant, remaining at the level of $\lesssim 0.3$ mas. However, the maximum change in $\Delta d_{SS}$ induced by the ring model still differs from other models by more than 0.8 km, which is much larger than the current observational accuracy of about 0.025 km for the geocentric Saturn distance.

In summary, when predicting Saturn's heliocentric distance, compared to the observed Plutinos, the arc model provides a better representation, as it results in a discrepancy of about 1 km smaller than that of the ring model. Given that the observational accuracy of the geocentric positions of Saturn is as small as $\sim0.025$ km, we suggest that future ephemeris computations, when modeling Plutinos, take into account their unique distribution driven by Neptune's 2:3 MMR, rather than assuming a homogeneous ring model.

\begin{table}
    \centering
    \begin{tabular}{l|c|c|c|c}
        \hline
       & $\Delta \lambda_S$ (mas) & $\Delta \phi_S$ (mas) & $\Delta d_{SN}$ (km) \\
           \hline
        \textbf{50 yr}\\
        \hline
       ring model             & -0.892 & 0.0082  &  0.467    \\
       arc model~($e=0.25$)   & -1.726 & 0.0382  &  1.507  \\
       DISpa Plutino samples  & -2.285 & 0.0673  & 2.182 \\
       L7 synthetic model & -1.593 & 0.0369 & 1.549 \\
       \hline
          \textbf{15 yr}\\
        \hline
       ring model             & -0.287 & 0.0039  &  0.467    \\
       arc model~($e=0.25$)   & -0.324 &  0.0056 &  1.343  \\
       DISpa Plutino samples  & -0.591 &  0.0052 & 2.160 \\
       L7 synthetic model & -0.251 & 0.0058 & 1.328 \\
            \hline
    \end{tabular}
    \caption{Similar to Table \ref{table:2}, but for Saturn's positions over 50 yr and 15 yr of evolution.}
    \label{table:3}
\end{table}

\section{Conclusions and discussion}
\label{sec:conclu}

This paper is devoted to developing an appropriate dynamical model to represent the Plutinos, which are KBOs trapped in the 2:3 MMR with Neptune, and to estimate their total perturbation by including the peculiar features of this MMR. Previously, the total perturbation of Plutinos was simulated using a one-dimensional, homogeneous ring model. However, since the 2:3 MMR involves three stable equilibrium points, Plutinos are expected to exhibit a unique spatial distribution. Consequently, the ring model may introduce a certain level of inaccuracy.

We analysed the potential azimuthal distribution of Plutinos resulting from the libration of the resonant angle $\sigma$ of Neptune's 2:3 MMR. Since $\sigma$ does not complete a full circulation from 0 to $360^{\circ}$ (i.e. the resonant amplitude $A<180^{\circ}$), the mean longitudes $\lambda$ of Plutinos relative to Neptune will be confined to three overlapping arcs. The length of each arc is equal and determined by the maximum resonant amplitude $A_{max}$, with $120^{\circ}$ serving as an upper limit for stable Plutinos. To capture this azimuthal distribution, we introduced a new three-arc model to represent the perturbations induced by Plutinos. Each arc is represented by a number of point masses, and a minimum number density of $6000/2\pi$ per radian in the azimuth has been deduced for effectively representing their gravitational perturbations. Accordingly, our three-arc model comprises a total of 9000 point masses, each arc possessing an equal mass.

We then assessed the total perturbation of Plutinos by calculating the change in the Sun-Neptune distance, $\Delta d_{SN}$, between the perturbed (including Plutinos) and unperturbed (excluding Plutinos) solar systems over the 100-yr period. For comparison, we employed both the ring and arc models to simulate the perturbations of Plutinos. Given Plutinos having eccentricities of $e=0.05$, our three-arc model with $A_{max}=10^{\circ}$ yielded a peak $\Delta d_{SN}$ of 89 km. In contrast, the reference case of a homogeneous ring resulted in a significantly larger $\Delta d_{SN}$, reaching approximately 154 km. This discrepancy is anticipated since a small $A_{max}$ can lead to a severe asymmetry in the azimuthal distribution of Plutinos. Nevertheless, this asymmetry noticeably decreases when considering the largest $A_{max}=120^{\circ}$, with the relative difference in $\Delta d_{SN}$ between the arc and ring models being about $10\%$ over the 100-yr period.

As a matter of fact, the azimuthal distribution of Plutino arcs is governed by $A_{max}$, while their radial distribution depends on $e$. It is interesting to discover that, with increasing $e$ from 0.05 to 0.1-0.2, the difference in $\Delta d_{SN}$ between the arc and ring model becomes more prominent. Notably, for the largest $e \gtrsim 0.25$, we emphasize two key findings: (1) the perturbation caused by Plutinos on the Sun-Neptune distance could consistently remain small for any  $A_{max}$, with the resulting $\Delta d_{SN}$ being only 10-20 km within the considered 100-yr period. (2) The $\Delta d_{SN}$ value obtained in the arc model can deviate by as much as 170 km from that in the ring model.

To summarize the above results, we have provided a concise analytic expression in Eq. (\ref{eq:AeMt}) to estimate the change in the Sun-Neptune distance caused by the perturbations of Plutinos. This expression takes into account the resonant amplitudes, eccentricities, and total mass of Plutinos. 

Subsequently, we further verified the applicability of our three-arc model by comparing it to the perturbations induced by observed Plutinos. After reducing observational bias by simulating the realistic spatial distribution of the Plutinos, the results show that the arc model can mimic Plutinos very well, especially in the first 50 yr of evolution, characterized by a different in $\Delta d_{SN}$ as small as $<2$ km. Considering that the observational accuracy of Saturn positions is much higher and could better constrain the Plutino distribution, we then calculated the change in the Sun-Saturn distance $\Delta d_{SS}$ induced by different perturbation models. Taking the perturbation of observed Plutinos as the standard, we find that the difference in $\Delta d_{SS}$ for the arc model is on the order of 1 km, while for the ring model it is about twice as large, at approximately 2 km. Given the $\sim0.025$ km observational accuracy of the geocentric positions of Saturn, we propose that our arc model is more appropriate for representing the observed Plutinos. In addition, we re-examined the comparison between the continuous and discrete ring models and found that the difference in $\Delta d_{SS}$ between these two models is only about 0.03 km. This discrepancy is notably smaller than the differences between the considered models. Specifically, the ring model consistently differs from other models by more than 1 km in $\Delta d_{SS}$. Thus, we have confirmed that the discrete model provides an adequate approximation of the continuous one.

As a supplement, we also considered a separate population of point-mass Plutinos from the L7 synthetic model \citep{L7Kavelaars2009,L7Petit2011,L7Gladman2012}. When comparing the perturbations induced by the L7 Plutinos on both Neptune's and Saturn's positions, the differences observed in the arc model are consistently much smaller than those in the ring model. This may further emphasize the importance of incorporating the arc model into planetary ephemerides rather than the ring model.

Although the obtained results suggest that our three-arc model is more appropriate as it takes into account the $A$ and $e$ distributions of Plutinos, there are still several aspects that can be further refined:

(1) The arc model could be somewhat rough due to the limited observations of observed Plutinos. In our theoretical study, although the arcs are confined to azimuthal ranges narrower than $0^{\circ}$ to $360^{\circ}$, they are indeed evenly distributed within these specific ranges. We identified that currently only about 460 Plutinos have been observed, while there may be as many as 9000 Plutinos with diameters larger than 100 km \citep{alexandersen2016carefully}. Although we tried to reduce observational bias, a more intrinsic arc model could be constructed for the Plutino population from future observations. For example, the Plutinos may be sparse in certain parts of the arcs while being dense in others. This potential inhomogeneity in the distribution of Plutinos would further diminish the applicability of the ring model. To overcome this issue, we may develop our arc model by considering a series of sub-arcs, each with different $A_{max}$ values, embedded in the three main arcs.

(2) We considered the point masses in the arc model with orbital inclinations of $i=0$. As motioned in Sect. \ref{sec:intro}, this simplicity is reasonable for our initial study because the 2:3 MMR is an eccentricity-type resonance, and its feature is controlled by eccentricity rather than inclination. However, observations show that a significant fraction of observed Plutinos has $i\sim5^{\circ}$-$35^{\circ}$. It should be noted that, in Sect. \ref{sec:real}, when evaluating the influence on either the Sun-Neptune or Sun-Saturn distance, the point-mass Plutinos are not distributed in the plane, but have an inclination distribution similar to the observed Plutinos. Given the inclined Plutinos, the planer arc model can still fit their perturbations well. Nevertheless, a more advanced arc model could be developed for higher accuracy, which would involve multiple layers, designed to accommodate various $i$ values. Within each layer, a one-dimensional arc model would be implemented.

In recent years, several projects for surveying the KBOs have been proposed, such as the ecliptic Deep Drilling Field by the Large Synoptic Survey Telescope (LSST) \citep{jone16, trilling2018deep}. The discovery of a larger number of small and faint KBOs is anticipated, enabling us to impose even tighter constraints on the spatial distribution and the total mass of Plutinos. Consequently, a more refined KBO model can further be developed.

Finally, we note that the positions of Neptune are mainly obtained through astronomical observations. Due to limitations posed by the Earth's atmosphere and the accuracy of star catalogs, Neptune's positional accuracy is currently at the level of a few thousand kilometers \citep{folkner2014planetary,pitjeva2018mass,Stone_2001}. While the difference in computed positions of Neptune between the ring and arc models is a few hundred kilometers, which appears smaller than the observational uncertainty. But, it's crucial to consider that the perturbation of Plutinos on Neptune's position is at a similar magnitude, around one hundred kilometers. Therefore, if the improvement of the ephemeris is to be carried out by the individual population of Plutinos, it may be necessary to adopt the arc model to represent their total perturbation. The improvement of accuracy for the positions of other planets (e.g. Saturn) may also leads to such a need. Moreover, as introduced in Sect. 1, there are several other resonant populations among the KBOs, with the number of each population as large as that of the Plutinos. Our multiple-arc model can be easily generated for these resonant KBOs. And, the combined gravitational effect of all the resonant KBOs should be more noticeable. We anticipate future space missions, such as NASA's Uranus Orbiter and Probe mission, to greatly improve the positional accuracy of Neptune. After the observational improvement, the contribution of our arc model in developing the ephemeris calculation for resonant KBOs may lead to more accurate predictions of Neptune's positions.

\section*{Acknowledgements}
    
This work was supported by the National Natural Science Foundation of China (Nos. 12473061, 11973027, 11933001, 12150009), and National Key R\&D Program of China (2019YFA0706601). We would also like to express our sincere thanks to the anonymous referee for the valuable comments.

\section*{Data Availability}

The data underlying this article are available in the article and in its online supplementary material.




\bibliographystyle{mnras}
\bibliography{reference} 

\begin{thebibliography}{}
\makeatletter
\relax
\def\mn@urlcharsother{\let\do\@makeother \do\$\do\&\do\#\do\^\do\_\do\%\do\~}
\def\mn@doi{\begingroup\mn@urlcharsother \@ifnextchar [ {\mn@doi@} {\mn@doi@[]}}
\def\mn@doi@[#1]#2{\def\@tempa{#1}\ifx\@tempa\@empty \href {http://dx.doi.org/#2} {doi:#2}\else \href {http://dx.doi.org/#2} {#1}\fi \endgroup}
\def\mn@eprint#1#2{\mn@eprint@#1:#2::\@nil}
\def\mn@eprint@arXiv#1{\href {http://arxiv.org/abs/#1} {{\tt arXiv:#1}}}
\def\mn@eprint@dblp#1{\href {http://dblp.uni-trier.de/rec/bibtex/#1.xml} {dblp:#1}}
\def\mn@eprint@#1:#2:#3:#4\@nil{\def\@tempa {#1}\def\@tempb {#2}\def\@tempc {#3}\ifx \@tempc \@empty \let \@tempc \@tempb \let \@tempb \@tempa \fi \ifx \@tempb \@empty \def\@tempb {arXiv}\fi \@ifundefined {mn@eprint@\@tempb}{\@tempb:\@tempc}{\expandafter \expandafter \csname mn@eprint@\@tempb\endcsname \expandafter{\@tempc}}}

\bibitem[\protect\citeauthoryear{Alexandersen, Gladman, Kavelaars, Petit, Gwyn, Shankman  \& Pike}{Alexandersen et~al.}{2016}]{alexandersen2016carefully}
Alexandersen M.,  Gladman B.,  Kavelaars J.,  Petit J.-M.,  Gwyn S.~D.,  Shankman C.~J.,   Pike R.~E.,  2016, AJ, 152, 111

\bibitem[\protect\citeauthoryear{Bannister et~al.,}{Bannister et~al.}{2018}]{bannister2018ossos}
Bannister M.~T.,  et~al., 2018, ApJS, 236, 18

\bibitem[\protect\citeauthoryear{Bernardinelli et~al.,}{Bernardinelli et~al.}{2020}]{Bernardinelli_2020}
Bernardinelli P.~H.,  et~al., 2020, ApJS, 247, 32

\bibitem[\protect\citeauthoryear{Brown}{Brown}{2013}]{brown2013density}
Brown M.~E.,  2013, AJ, 778, L34

\bibitem[\protect\citeauthoryear{Brown \& Schaller}{Brown \& Schaller}{2007}]{brown2007mass}
Brown M.~E.,  Schaller E.~L.,  2007, Science, 316, 1585

\bibitem[\protect\citeauthoryear{Brown, Ragozzine, Stansberry  \& Fraser}{Brown et~al.}{2010}]{brown2010size}
Brown M.~E.,  Ragozzine D.,  Stansberry J.,   Fraser W.~C.,  2010, AJ, 139, 2700

\bibitem[\protect\citeauthoryear{Brozovi{\'c}, Showalter, Jacobson  \& Buie}{Brozovi{\'c} et~al.}{2015}]{brozovic2015orbits}
Brozovi{\'c} M.,  Showalter M.~R.,  Jacobson R.~A.,   Buie M.~W.,  2015, Icarus, 246, 317

\bibitem[\protect\citeauthoryear{Chen, Gladman, Volk, Murray-Clay  et~al.}{Chen et~al.}{2019}]{chenYT2019}
Chen Y.-T.,  Gladman B.,  Volk K.,  Murray-Clay R.,   et~al., 2019, AJ, 158, 214

\bibitem[\protect\citeauthoryear{Chiang \& Jordan}{Chiang \& Jordan}{2002}]{Chiang2002}
Chiang E.~I.,  Jordan A.~B.,  2002, AJ, 124, 3430

\bibitem[\protect\citeauthoryear{{Cohen} \& {Hubbard}}{{Cohen} \& {Hubbard}}{1965}]{1965AJCohen}
{Cohen} C.~J.,  {Hubbard} E.~C.,  1965, \mn@doi [AJ] {10.1086/109674}, \href {https://ui.adsabs.harvard.edu/abs/1965AJ.....70...10C} {70, 10}

\bibitem[\protect\citeauthoryear{Di~Ruscio, Fienga, Durante, Iess, Laskar  \& Gastineau}{Di~Ruscio et~al.}{2020}]{di2020analysis}
Di~Ruscio A.,  Fienga A.,  Durante D.,  Iess L.,  Laskar J.,   Gastineau M.,  2020, A\&A, 640, A7

\bibitem[\protect\citeauthoryear{Fernández, Gallardo  \& Brunini}{Fernández et~al.}{2004}]{FERNANDEZ2004372}
Fernández J.~A.,  Gallardo T.,   Brunini A.,  2004, \mn@doi [Icarus] {https://doi.org/10.1016/j.icarus.2004.07.023}, 172, 372

\bibitem[\protect\citeauthoryear{Fienga, Manche, Laskar  \& Gastineau}{Fienga et~al.}{2008}]{fienga2008inpop06}
Fienga A.,  Manche H.,  Laskar J.,   Gastineau M.,  2008, A\&A, 477, 315

\bibitem[\protect\citeauthoryear{Fienga et~al.,}{Fienga et~al.}{2020}]{fienga2020inpop}
Fienga A.,  et~al., 2020, in {Bizouard} C.,  ed., Astrometry, Earth Rotation, and Reference Systems in the GAIA era. pp 293--297

\bibitem[\protect\citeauthoryear{{Fienga}, {Deram}, {Di Ruscio}, {Viswanathan}, {Camargo}, {Bernus}, {Gastineau}  \& {Laskar}}{{Fienga} et~al.}{2021}]{fienga2021}
{Fienga} A.,  {Deram} P.,  {Di Ruscio} A.,  {Viswanathan} V.,  {Camargo} J.~I.~B.,  {Bernus} L.,  {Gastineau} M.,   {Laskar} J.,  2021, Notes Scientifiques et Techniques de l'Institut de Mecanique Celeste, \href {https://ui.adsabs.harvard.edu/abs/2021NSTIM.110.....F} {110}

\bibitem[\protect\citeauthoryear{Fienga, Bigot, Mary, Deram, Di~Ruscio, Bernus, Gastineau  \& Laskar}{Fienga et~al.}{2022}]{fienga2022evolution}
Fienga A.,  Bigot L.,  Mary D.,  Deram P.,  Di~Ruscio A.,  Bernus L.,  Gastineau M.,   Laskar J.,  2022, IAU Symp., 364, 31

\bibitem[\protect\citeauthoryear{Folkner, Williams, Boggs, Park  \& Kuchynka}{Folkner et~al.}{2014}]{folkner2014planetary}
Folkner W.~M.,  Williams J.~G.,  Boggs D.~H.,  Park R.~S.,   Kuchynka P.,  2014, IPN Progress Report, \href {https://ui.adsabs.harvard.edu/abs/2014IPNPR.196C...1F} {42-196, 1}

\bibitem[\protect\citeauthoryear{Fraser, Batygin, Brown  \& Bouchez}{Fraser et~al.}{2013}]{fraser2013mass}
Fraser W.~C.,  Batygin K.,  Brown M.~E.,   Bouchez A.,  2013, Icarus, 222, 357

\bibitem[\protect\citeauthoryear{{Gladman}, {Marsden}  \& {Vanlaerhoven}}{{Gladman} et~al.}{2008}]{gladman2008}
{Gladman} B.,  {Marsden} B.~G.,   {Vanlaerhoven} C.,  2008, in {Barucci} M.~A.,  {Boehnhardt} H.,  {Cruikshank} D.~P.,  {Morbidelli} A.,   {Dotson} R.,  eds, , The Solar System Beyond Neptune.
pp 43--57

\bibitem[\protect\citeauthoryear{{Gladman} et~al.,}{{Gladman} et~al.}{2012}]{L7Gladman2012}
{Gladman} B.,  et~al., 2012, \mn@doi [\aj] {10.1088/0004-6256/144/1/23}, \href {https://ui.adsabs.harvard.edu/abs/2012AJ....144...23G} {144, 23}

\bibitem[\protect\citeauthoryear{Grav et~al.,}{Grav et~al.}{2011}]{grav2011wise}
Grav T.,  et~al., 2011, AJ, 742, 40

\bibitem[\protect\citeauthoryear{Grav, Mainzer, Bauer, Masiero  \& Nugent}{Grav et~al.}{2012}]{grav2012wise}
Grav T.,  Mainzer A.~K.,  Bauer J.~M.,  Masiero J.~R.,   Nugent C.~R.,  2012, AJ, 759, 49

\bibitem[\protect\citeauthoryear{Grundy et~al.,}{Grundy et~al.}{2015}]{grundy2015mutual}
Grundy W.~M.,  et~al., 2015, Icarus, 257, 130

\bibitem[\protect\citeauthoryear{Huang \& Zhou}{Huang \& Zhou}{1993}]{tian1993adams}
Huang T.-Y.,  Zhou Q.-L.,  1993, ChA\&A, 17, 205

\bibitem[\protect\citeauthoryear{Jewitt, Luu  \& Trujillo}{Jewitt et~al.}{1998}]{jewitt1998large}
Jewitt D.,  Luu J.,   Trujillo C.,  1998, AJ, 115, 2125

\bibitem[\protect\citeauthoryear{Jewitt, Sheppard  \& Porco}{Jewitt et~al.}{2004}]{jewi04}
Jewitt D.~C.,  Sheppard S.,   Porco C.,  2004, in Bagenal~F. D.~T.,  McKinnon W.~E.,  eds, , Jupiter: The Planet, Satellites and Magnetosphere.
Cambridge: Cambridge Univ. Press, p.~263

\bibitem[\protect\citeauthoryear{{Jones}, {Fomalont}, {Dhawan}, {Romney}, {Folkner}, {Lanyi}, {Border}  \& {Jacobson}}{{Jones} et~al.}{2011}]{Jones2011}
{Jones} D.~L.,  {Fomalont} E.,  {Dhawan} V.,  {Romney} J.,  {Folkner} W.~M.,  {Lanyi} G.,  {Border} J.,   {Jacobson} R.~A.,  2011, \mn@doi [\aj] {10.1088/0004-6256/141/2/29}, \href {https://ui.adsabs.harvard.edu/abs/2011AJ....141...29J} {141, 29}

\bibitem[\protect\citeauthoryear{Jones, Juri\'c  \& Ivezi\'c}{Jones et~al.}{2015}]{jone16}
Jones R.~L.,  Juri\'c M.,   Ivezi\'c v.,  2015, in , Vol.~318, Proceedings of the International Astronomical Union.
Cambridge Univ. Press, p.~282

\bibitem[\protect\citeauthoryear{{Kavelaars} et~al.,}{{Kavelaars} et~al.}{2009}]{L7Kavelaars2009}
{Kavelaars} J.~J.,  et~al., 2009, \mn@doi [\aj] {10.1088/0004-6256/137/6/4917}, \href {https://ui.adsabs.harvard.edu/abs/2009AJ....137.4917K} {137, 4917}

\bibitem[\protect\citeauthoryear{Kiss et~al.,}{Kiss et~al.}{2019}]{kiss2019mass}
Kiss C.,  et~al., 2019, Icarus, 334, 3

\bibitem[\protect\citeauthoryear{Krasinsky, Pitjeva, Vasilyev  \& Yagudina}{Krasinsky et~al.}{2002}]{KRASINSKY200298}
Krasinsky G.~A.,  Pitjeva E.~V.,  Vasilyev M.~V.,   Yagudina E.~I.,  2002, \mn@doi [Icarus] {https://doi.org/10.1006/icar.2002.6837}, 158, 98

\bibitem[\protect\citeauthoryear{Kuchynka, Laskar, Fienga  \& Manche1}{Kuchynka et~al.}{2010}]{Kuchynka2010}
Kuchynka P.,  Laskar J.,  Fienga A.,   Manche1 H.,  2010, A\&A, 514, A96

\bibitem[\protect\citeauthoryear{Lawler et~al.,}{Lawler et~al.}{2018}]{lawler2018ossos}
Lawler S.~M.,  et~al., 2018, AJ, 155, 197

\bibitem[\protect\citeauthoryear{Li \& Sun}{Li \& Sun}{2018}]{li2018constructing}
Li J.,  Sun Y.-S.,  2018, A\&A, 616, A70

\bibitem[\protect\citeauthoryear{Li \& Xia}{Li \& Xia}{2020}]{li2020Planet9}
Li J.,  Xia Z.~J.,  2020, A\&A, 637, A87

\bibitem[\protect\citeauthoryear{Li, Zhou  \& Sun}{Li et~al.}{2007}]{li2007origin}
Li J.,  Zhou L.-Y.,   Sun Y.-S.,  2007, A\&A, 464, 775

\bibitem[\protect\citeauthoryear{Li, Zhou  \& Sun}{Li et~al.}{2014}]{li2014study}
Li J.,  Zhou L.-Y.,   Sun Y.-S.,  2014, MNRAS, 437, 215

\bibitem[\protect\citeauthoryear{Li, Xia  \& Zhou}{Li et~al.}{2019}]{li2019calibration}
Li J.,  Xia Z.~J.,   Zhou L.,  2019, A\&A, 630, A68

\bibitem[\protect\citeauthoryear{Li, Lawler, Zhou  \& Sun}{Li et~al.}{2020}]{li2020study}
Li J.,  Lawler S.~M.,  Zhou L.-Y.,   Sun Y.-S.,  2020, MNRAS, 492, 3566

\bibitem[\protect\citeauthoryear{Li, Li, Xia  \& Georgakarakos}{Li et~al.}{2022}]{li2022machine}
Li X.,  Li J.,  Xia Z.~J.,   Georgakarakos N.,  2022, MNRAS, 511, 2218

\bibitem[\protect\citeauthoryear{Li, Lawler  \& Lei}{Li et~al.}{2023a}]{li2023study}
Li J.,  Lawler S.~M.,   Lei H.,  2023a, MNRAS, 523, 4841

\bibitem[\protect\citeauthoryear{Li, Xia, Yoshida, Georgakarakos  \& Li}{Li et~al.}{2023b}]{li2023JT}
Li J.,  Xia Z.~J.,  Yoshida F.,  Georgakarakos N.,   Li X.,  2023b, A\&A, 669, A68

\bibitem[\protect\citeauthoryear{Liu, Fienga  \& Yan}{Liu et~al.}{2022}]{liu2022ring}
Liu S.,  Fienga A.,   Yan J.,  2022, Icarus, 376, 114845

\bibitem[\protect\citeauthoryear{Murray \& Dermott}{Murray \& Dermott}{2000}]{murray_dermott_2000}
Murray C.~D.,  Dermott S.~F.,  2000, Solar System Dynamics.
Cambridge University Press, p. 225–273

\bibitem[\protect\citeauthoryear{{Nesvorn{\'y}}, {Bro{\v{z}}}  \& {Carruba}}{{Nesvorn{\'y}} et~al.}{2015}]{Nesv2015}
{Nesvorn{\'y}} D.,  {Bro{\v{z}}} M.,   {Carruba} V.,  2015, in , Asteroids IV.
Univ. Arizona Press, pp 297--321

\bibitem[\protect\citeauthoryear{Newhall, Standish  \& Williams}{Newhall et~al.}{1983}]{newhall1983102}
Newhall X.~X.,  Standish E.~M.,   Williams J.~G.,  1983, A\&A, 125, 150

\bibitem[\protect\citeauthoryear{Park, Folkner, Williams  \& Boggs}{Park et~al.}{2021}]{Park_2021}
Park R.~S.,  Folkner W.~M.,  Williams J.~G.,   Boggs D.~H.,  2021, \mn@doi [The Astronomical Journal] {10.3847/1538-3881/abd414}, 161, 105

\bibitem[\protect\citeauthoryear{{Petit} et~al.,}{{Petit} et~al.}{2011}]{L7Petit2011}
{Petit} J.~M.,  et~al., 2011, \mn@doi [\aj] {10.1088/0004-6256/142/4/131}, \href {https://ui.adsabs.harvard.edu/abs/2011AJ....142..131P} {142, 131}

\bibitem[\protect\citeauthoryear{Petit, Gladman, Kavelaars, Bannister, Alexandersen, Volk  \& Chen}{Petit et~al.}{2023}]{petit2023hot}
Petit J.-M.,  Gladman B.,  Kavelaars J.,  Bannister M.~T.,  Alexandersen M.,  Volk K.,   Chen Y.-T.,  2023, ApJL, 947, L4

\bibitem[\protect\citeauthoryear{Pitjeva}{Pitjeva}{2001}]{pitjeva2001modern}
Pitjeva E.~V.,  2001, Celest. Mech. Dyn. Astron., 80, 249

\bibitem[\protect\citeauthoryear{Pitjeva}{Pitjeva}{2010}]{pitjeva_2010}
Pitjeva E.~V.,  2010, \mn@doi [IAU Symp.] {10.1017/S1743921310001560}, 263, 93

\bibitem[\protect\citeauthoryear{Pitjeva}{Pitjeva}{2013}]{Pitjeva2013}
Pitjeva E.~V.,  2013, Solar System Research, 47, 386

\bibitem[\protect\citeauthoryear{Pitjeva \& Pitjev}{Pitjeva \& Pitjev}{2014}]{pitjeva2014development}
Pitjeva E.~V.,  Pitjev N.~P.,  2014, Celest. Mech. Dyn. Astron., 119, 237

\bibitem[\protect\citeauthoryear{Pitjeva \& Pitjev}{Pitjeva \& Pitjev}{2018a}]{pitjeva2018masses}
Pitjeva E.~V.,  Pitjev N.~P.,  2018a, Astronomy Letters, 44, 554

\bibitem[\protect\citeauthoryear{Pitjeva \& Pitjev}{Pitjeva \& Pitjev}{2018b}]{pitjeva2018mass}
Pitjeva E.~V.,  Pitjev N.~P.,  2018b, Celest. Mech. Dyn. Astron, 130, 1

\bibitem[\protect\citeauthoryear{Ragozzine \& Brown}{Ragozzine \& Brown}{2009}]{ragozzine2009orbits}
Ragozzine D.,  Brown M.~E.,  2009, AJ, 137, 4766

\bibitem[\protect\citeauthoryear{Standish}{Standish}{1998}]{standish1998jpl}
Standish E.~M.,  1998, Jet Propulsion Laboratory, IOM 312.F-98-048

\bibitem[\protect\citeauthoryear{Stansberry et~al.,}{Stansberry et~al.}{2012}]{stansberry2012physical}
Stansberry J.~A.,  et~al., 2012, Icarus, 219, 676

\bibitem[\protect\citeauthoryear{Stone}{Stone}{2001}]{Stone_2001}
Stone R.~C.,  2001, \mn@doi [AJ] {10.1086/323549}, 122, 2723

\bibitem[\protect\citeauthoryear{Szab{\'o}, Ivezi{\'c}, Juri{\'c}  \& Lupton}{Szab{\'o} et~al.}{2007}]{szabo2007properties}
Szab{\'o} G.~M.,  Ivezi{\'c} {\v{Z}}.,  Juri{\'c} M.,   Lupton R.,  2007, MNRAS, 377, 1393

\bibitem[\protect\citeauthoryear{Tian}{Tian}{2023}]{Tian2023}
Tian W.,  2023, Celest. Mech. Dyn. Astron., 135, 38

\bibitem[\protect\citeauthoryear{Trilling, Bannister, Fuentes, Gerdes, Mommert, Schwamb  \& Trujillo}{Trilling et~al.}{2018}]{trilling2018deep}
Trilling D.~E.,  Bannister M.,  Fuentes C.,  Gerdes D.,  Mommert M.,  Schwamb M.~E.,   Trujillo C.,  2018, arXiv preprint arXiv:1812.09705

\bibitem[\protect\citeauthoryear{{Trujillo} et~al.,}{{Trujillo} et~al.}{2022}]{2022DPS....5450101T}
{Trujillo} C.,  et~al., 2022, in AAS/Division for Planetary Sciences Meeting Abstracts. p. 501.01

\bibitem[\protect\citeauthoryear{Volk et~al.,}{Volk et~al.}{2016}]{volk2016ossos}
Volk K.,  et~al., 2016, AJ, 152, 23

\makeatother
\end{thebibliography}




\appendix

\section{Earth positions in the barycentric coordinate system}

The introduction of the Plutinos in planetary ephemerides will induce a drift in the solar system barycenter (SSB), which is defined by the centre of mass in each proposed model. Although the relative coordinates between two celestial bodies, such as Sun-planet or planet-planet, remain the same across different models \citep{pitjeva2014development}, it is also very interesting to provide a glimpse of the SSB-Earth positions.

\begin{figure}
\includegraphics[width=\columnwidth]{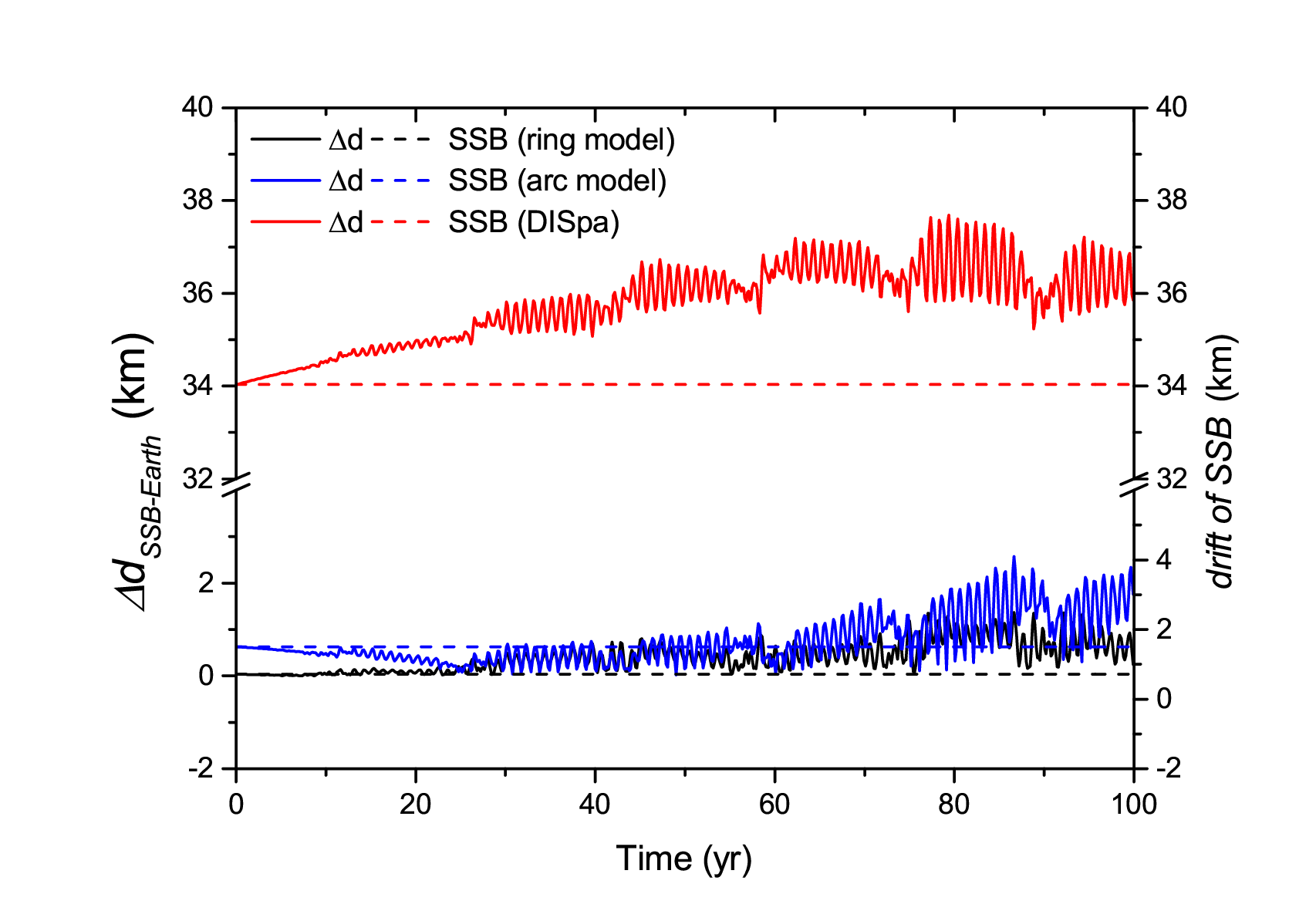}
\caption{Perturbations on the SSB-Earth distance induced by the arc model (blue solid curve), ring model (black solid curve) and the point-mass Plutinos corrected by the DISpa method (red solid curve). For reference, the drifts of the SSB induced by different models are shown by the horizontal dashed lines.}
  \label{fig:app1}
\end{figure}

In Fig. \ref{fig:app1}, we show the effect of three representative models on the SSB-Earth distance ($\Delta d_{SSB-Earth}$), represented by the solid curves. The amplitudes of the $\Delta d_{SSB-Earth}$ variations consistently range between 2 and 4 km. Nevertheless, one may notice that the $\Delta d_{SSB-Earth}$ values induced by the DISpa samples (in red) are about 30 km larger than those of the arc (blue) and ring (in black) models. This difference is due to the drift of the SSB in these models, as indicated by the horizontal dashed lines. It is worth noting again that the inclusion of all asteroids in the solar system can cause a displacement of the SSB by as much as 100 km \citep{li2019calibration}.

\begin{figure}
    \includegraphics[width=\columnwidth]{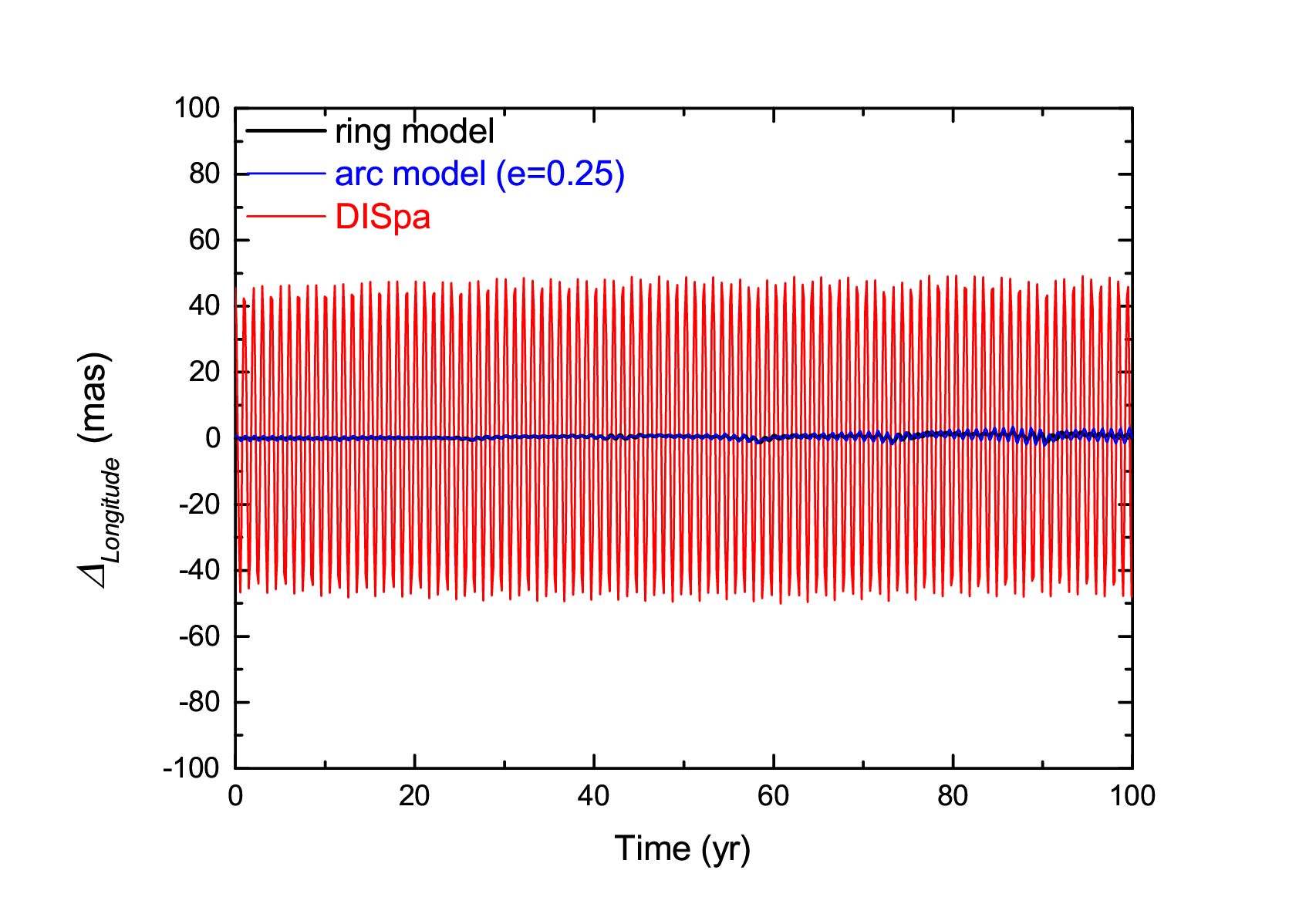}
    \includegraphics[width=\columnwidth]{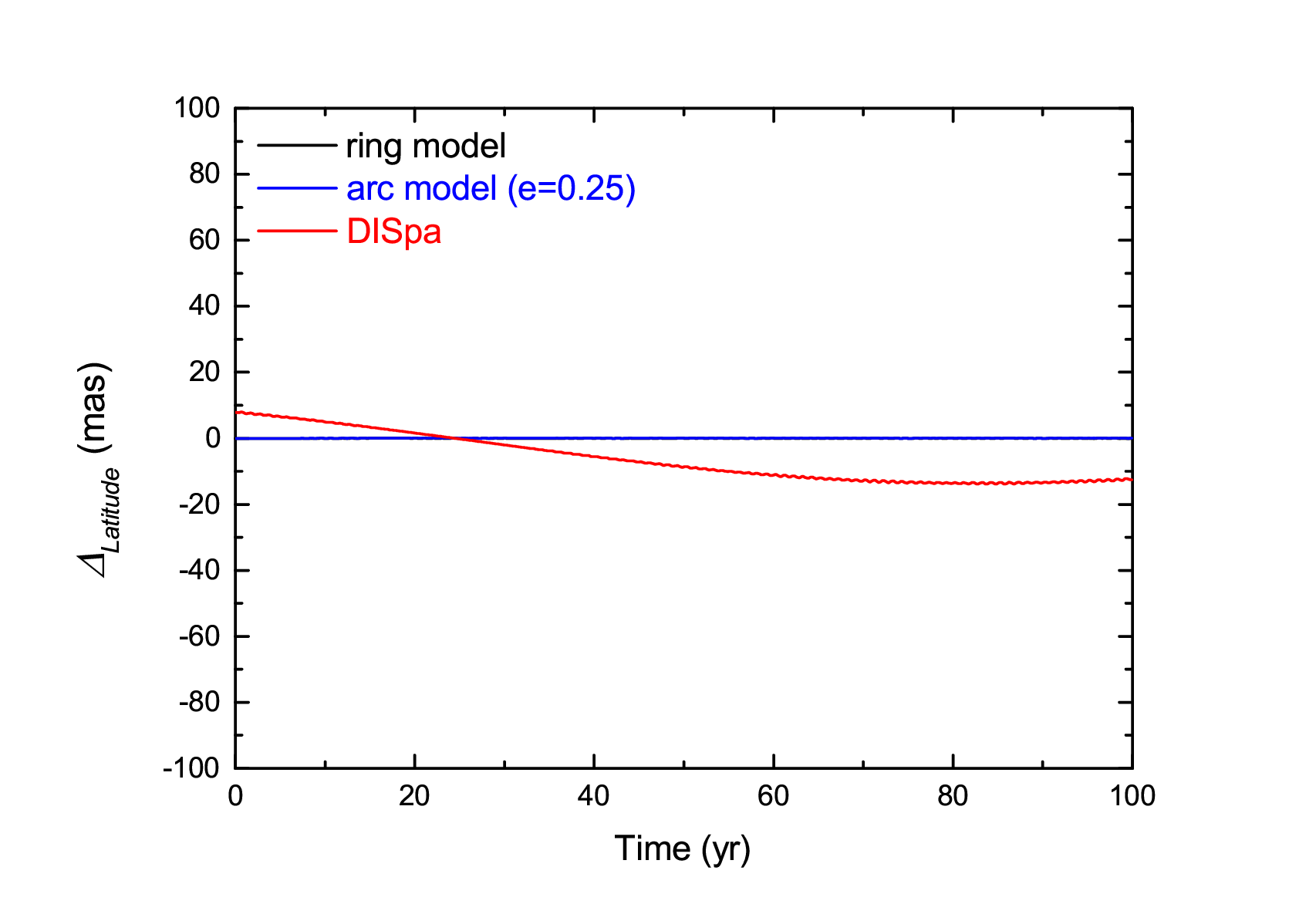}
\caption{Similar to Fig. \ref{fig:app1}, but showing the variations in Earth's longitude ($\Delta_{Longitude}$) and latitude ($\Delta_{Latitude}$) in the barycentric coordinate system.}
\label{fig:app2}
\end{figure}

With respect to Earth's longitude ($\Delta_{Longitude}$) and latitude ($\Delta_{Latitude}$) in the barycentric coordinate system, Fig. \ref{fig:app2} shows that the variations in these angular coordinates are less than 1 mas in both the arc and ring models. While for the point-mass Plutinos generated by the DISpa method, the maximum changes in $\Delta_{Longitude}$ and $\Delta_{Latitude}$ reach approximately 40 mas and 20 mas, respectively. We believe these relatively large changes can be attributed to the fact that the observational bias has not been completely eliminated, as discussed in the main text. According to the calculations in \citet{li2019calibration}, the contribution of the faintest and least massive asteroids can change the ascending node of the invariable plane of the solar system in the barycentric frame by about 40 mas. This seems to align with the results concerning the impact of observational bias.

\bsp	
\label{lastpage}
\end{document}